\documentclass[amsmath,amssymb,
pra,
]
{revtex4}
\usepackage{graphicx}
\usepackage{dcolumn}
\usepackage{bm}

\newcommand{\ket}[1]{\ensuremath{| #1 \rangle}}

\begin{document}

\title{Tunneling of ultracold atoms in time-independent potentials}
\author{Ennio Arimondo$^{1}$ and Sandro Wimberger$^{2}$}

\affiliation{$^1$INO-CNR and Dipartimento di Fisica Enrico Fermi, Universit\`{a} di Pisa, Largo Pontecorvo 3, I-56127 Pisa\\
$^2$Institut f\"ur Theoretische Physik, Universit\"at Heidelberg, Philosophenweg 19, D-69120 Heidelberg}

\date{\today}

\begin{abstract} 
We present theoretical as well as experimental results on  resonantly enhanced quantum tunneling of Bose-Einstein
condensates in optical lattices both in the linear case of single particle dynamics 
and in the presence of atom-atom interactions. Our results demonstrate the usefulness of
condensates in optical lattices for the dynamical control of tunneling and for simulating Hamiltonians
originally used for describing solid state phenomena.
\end{abstract}

\maketitle


\section{\label{sec1} Introduction}

\begin{figure}
\centering
\includegraphics[height=10cm]{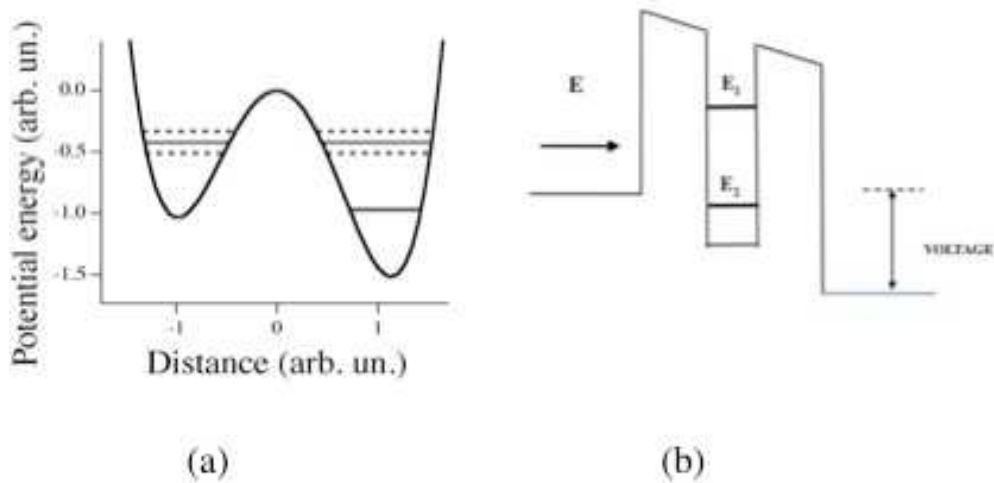}
\caption{(a) schematic representation of the energy levels within an asymmetric double well. The unperturbed energies within the left and right wells are indicated by the continuous lines. Because of the resonant tunneling  between the ground state in the left well and the first excited one in the right well, the asymmetric and antisymmetric states have energy indicated by the dashed lines. (b) schematic band diagram of a resonant-tunneling diode structure under a voltage bias between the incoming (left) and outgoing (right) regions.} 
\label{DoubleBarrier}
\end{figure}

 Tunneling as a quantum mechanical effect takes place in a classically forbidden region between two regions of classically allowed motion. While the term ``dynamical tunneling'' typically refers to tunneling of quantum states across dynamical barriers in classical phase space \cite{DavisHeller1981}, the original problem simply intended tunneling across a potential barrier. Both types of tunneling are addressed in this chapter, with major focus on situations in which external forces make the studied systems intrinsically time-dependent and allow for a dynamical control of tunneling through potential barriers or across band gaps which are dynamically explored by the system. \\
\indent A standard example of tunneling across static barriers is the motion in a double well potential. The two potential wells are separated by a potential barrier which is impenetrable for a low-energy classical particle. The quantum mechanical solution  shows that the wave packet initially localized in one of the wells performs oscillations between the two classically allowed region.  Tunneling takes place between two levels nearly degenerate in energy, and in most cases the investigated tunneling takes place between the lowest energy states, for instance of a double well. However, in a potential configuration as the asymmetric double well shown in Fig.~\ref{DoubleBarrier}(a),  an energy matching between a ground state on one side and an excited state on the other side leads to a tunneling between those states resonantly enhanced by the energy matching.   In the resonantly enhanced tunneling (RET)  the probability for  the quantum tunneling of a particle between two potential wells is increased when the energies of the initial and final states of the process coincide. In the one-dimensional double potential barrier of Fig.~\ref{DoubleBarrier}(b), the narrow central potential
well has weakly-quantized (or quasi-stationary) bound states, of which the energies are denoted by $E_1$ and $E_2$ in Fig.~\ref{DoubleBarrier}. If the energy $E$ of electrons incident on the barrier coincides with these energies, the electrons may tunnel through both barriers without any attenuation. 
The transmission coefficient reaches unity at the electron energy $E = E_1$ or
$E=E_2$.  It is interesting that while the transmission coefficient of a potential barrier is always lower than one, two barriers in a row can be completely transparent for certain energies of the incident particle. \\
\indent In the early 1970s, Tsu, Esaki, and Chang computed the two terminal current-voltage characteristics of a finite superlattice, and predicted that RET to be observed not only in the transmission coefficient but also in the current-voltage characteristic~\cite{Tsu1973,Chang1974}. Resonant tunneling also occurs in potential profiles with more than two barriers. Technical advances led to the observation of negative differential conductance at terahertz frequencies and triggered a considerable research effort to study tunneling through multi-barrier structures. 
Owing to the fundamental nature of this effect and the practical
interest~\cite{Chang1991}, in the last few years much progress has
been made in constructing solid state systems such as
superlattices~\cite{Esaki1986,Leo2003,Glutsch2004},
quantum wells~\cite{Wagner1993}, and waveguide arrays
\cite{Rosam2003} which enable the controlled observation and application of RET.
The potential profiles required for resonant tunneling and realized in semiconductor system using heterojunctions allowed the manufacture of resonant tunneling diodes. These devices have important applications such as in high-frequency signal generation and multi-valued data storage, as reviewed in~\cite{Mizuta1995}. \\
\indent In the last decade, the experimental techniques used in atom and
quantum optics have made it possible to control the external and
internal degrees of freedoms of ultracold atoms with a very high
degree of precision. Thus, ultracold bosons or fermions loaded
into the periodic optical potential created by interfering laser beams (double-well, 
lattices and superlattices) are optimal realizations of quantum mechanical processes and phenomena proposed 
and studied in other contexts of solid-state physics. Ultracold atoms and Bose-Einstein condensates, 
for instance, have been used to simulate phenomena such as Bloch oscillations in tilted periodic
potentials \cite{Peik1996,Raizen1997,Roati2004,Morsch2006,Gustavsson2008} and to study quantum phase 
transitions driven by atom-atom interactions \cite{Bloch2008}. \\
\indent RET-like effects have been observed in a number of experiments to date. In ref. \cite{Teo2002}, resonant tunneling
was observed for cold atoms trapped by an optical lattice when an
applied magnetic field produced a Zeeman splitting of the energy
levels.  Resonant tunneling has been observed in a Mott insulator within an optical lattice, 
where a finite amount of energy given by the on-site interaction energy is
required to create a particle-hole excitation \cite{Greiner2002}. Tunneling of the
atoms is therefore suppressed.  If the lattice potential is tilted
by application of a  potential gradient, RET is allowed whenever
the energy difference between neighboring lattice sites due to
the potential gradient matches the on-site interaction energy. This 
RET  control in a Mott insulator allowed F\"olling {\it et al.} to observe 
a second-order coherence, i.e. a two-atom RET~\cite{Foelling2007}. \\
\indent Most of the quantum transport phenomena investigated
with Bose-Einstein condensates within periodic optical lattices
focused on the atomic motion in the ground state band of the
periodic lattice. Only a few experiments examined the quantum
transport associated with interband transitions ``vertical'' in
the energy space. Interband transitions were induced by additional
electromagnetic  fields, as in the case of the spectroscopy of
Wannier-Stark levels \cite{Wilkinson1996}, or by quantum tunneling between
the bands. Tunneling between otherwise uncoupled energy bands
occurs  when the bands are coupled by an additional force, which
can be a static Stark force (tilting the otherwise periodic
lattice) \cite{Morsch2006}, or also by strong atom-atom interactions
 as observed for fermions in \cite{Koehl2005} and discussed 
for bosons in \cite{Lee2007}. The quantum tunneling between the ground and the
first excited band is particularly pronounced in the presence of
degeneracies of the single-well energy levels within the optical
lattice leading to RET.   In~\cite{Sias2007,Zenesini2008} such a type of RET was investigated 
 for a Bose-Einstein condensate in a one-dimensional optical lattice, which allows for a
high level of control on the  potential depth and the lattice  tilt. 
 Those experimental investigations concentrated on the regime of parameters for which the tilting force -- at RET conditions equal to the energy difference between neighboring wells -- dominated the dynamics of the condensate. The 
RET tunneling of the ground band and the first two excited energy bands  
were measured in a wide range of experimental conditions.  In addition the RET process is modified by
the atom-atom interactions, bringing new physics to the quantum tunneling. \\
\indent This chapter is organized as follows. Section II sets the stage discussing optical lattices and giving the necessary background.
While Section III reports on RET in closed two and three well systems, Section IV focuses on our main subject, the control
of tunneling by RET in open quantum systems. This Section reports on our experimental data
in the linear tunneling regime, i.e. in the absence of atom-atom interactions, as well as on interaction induced effects.
In Section V a model for many-body tunneling is introduced,  before we summarize the recent advances concerning RET in Section \ref{sec6}.

\begin{figure}
\centering
\includegraphics[height=8cm]{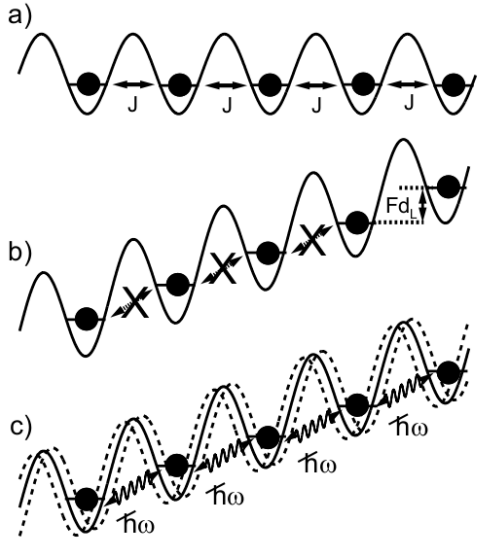}
\caption{(a) In an optical lattice without additional external forces, the ground-state levels are resonantly coupled, leading to a tunneling energy $J$. (b) When a linear
potential is applied, e.g. by applying a force $F$, the levels
are shifted out of resonance and tunneling is suppressed (Wannier-Stark localization). (c) If
an additional potential energy oscillating at an appropriate frequency $\omega$ is applied, the levels can again be coupled through photons of
energy $\hbar \omega$ and tunneling is partially restored.} 
\label{TunnelingOL}
\end{figure}
\begin{figure}
\centering
\includegraphics[height=7cm]{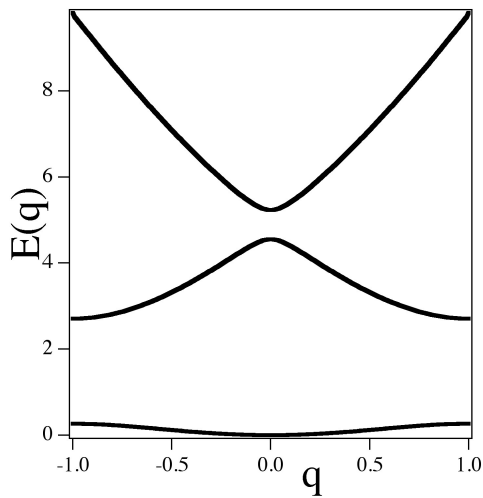}
\caption{Plot of the energies for the energy bands $E_{\rm n}(q)$ versus quasimomentum $q$ for an optical lattice with optical depth $V_0=4E_{\rm rec}$.} 
\label{bands}
\end{figure}

\section{\label{sec2} Optical lattices}
\begin{figure}
\centering
\includegraphics[height=6cm]{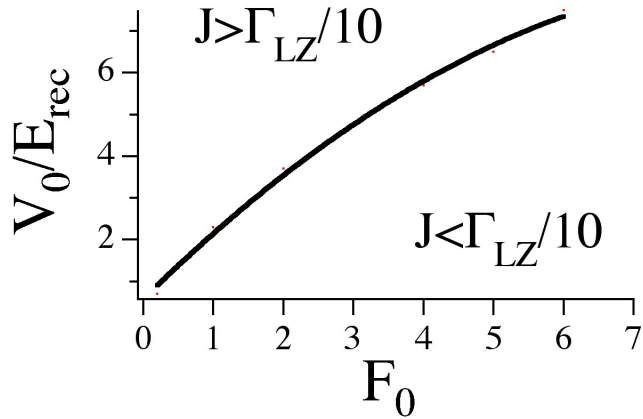}
\caption{Plots of the line in the space of the optical lattice depth $V_0$, in $E_{\rm rec}$ units, and the dimensionless force $F_0$ dividing the upper (lower) region where the interwell tunneling is ten times larger (smaller) than the Landau-Zener tunneling to the upper band.} \label{LandauZenerLimit}
\end{figure}

\indent The investigations of tunneling for cold/ultracold atoms (Bose-Einstein condensates or Fermi degenerate gases)  are based on the use of optical lattices~\cite{Cristiani2002,Morsch2006}. For a 1D optical lattice a standing wave is created  by the interference of two linearly polarized
traveling waves counter-propagating along the $x$-axis with  frequency $\omega_{\rm L}$ and wave-vector $\lambda_{\rm L}$. The amplitude of the generated electric field is ${\cal{E}}(r, t) = 2{\cal{E}}_{\rm 0} \sin(\omega_{\rm L}t) \sin(\frac{2\pi x}{\lambda_{\rm L}})$. When the laser detuning from the atomic transition is large enough to neglect the excited state spontaneous
emission decay, the atom experiences a periodically varying
conservative potential
\begin{equation}
V_{\rm ol} (x) =V_0 \sin^2\left(\frac{\pi x}{d_L}\right),
\label{potential}
\end{equation}
schematically represented in Fig.~\ref{TunnelingOL}(a). The amplitude $V_0$ depends on the laser detuning from the atomic transition and on the  square of the $\cal{E}_{\rm 0}$ electric field amplitude \cite{Grimm2000}. The periodic potential has a spacing $d_{\rm L} = \lambda_{\rm L}/2$. This potential derives from the quantum mechanical  interaction between atom and 
optical lattice photons. Therefore the  lattice quantities  are  linked to the recoil momentum $p_{\rm rec} = 2 \pi\hbar/\lambda_{\rm L}$  acquired by an atom after the absorption or the emission of one photon. $V_{\rm 0}$ will be expressed in units of $E_{\rm rec}$, the recoil energy acquired by an atom having mass $M$ following one photon exchange
\begin{equation}
E_{\rm rec} =\frac{h^2}{2M \lambda_{\rm L}^2}.
\label{recoilenergy}
\end{equation}
Neglecting the atom-atom interactions in a Bose-Einstein condensate, our 1D system is described by the following Hamiltonian:
\begin{equation}
H = -\frac{\hbar^2}{2M}\frac{d^2}{dx^2} + V_0 \sin^2\left(\frac{\pi x}{d_L}\right).
\label{freeatoms}
\end{equation}
For  this periodic potential  the associated single-particle
eigenstates in the lowest band are Bloch plane waves with quasimomentum
$q$. The energies $E_{\rm n}(q)$ of the Bloch waves for the lowest bands $n=1,2,3$ are plotted in Fig. \ref{bands} versus quasimomentum. Ultracold atoms are loaded into the ground state band having a minimum gap $\Delta$ at the edge of the Brillouin zone. The atomic evolution within that band or the excitation to a higher band is typically investigated. \\
\indent If a force $F$ is applied to the atom, as schematized in Fig.~\ref{TunnelingOL}(b), the following Hamiltonian describes the atomic evolution neglecting for a moment atom-atom interactions in a Bose-Einstein condensate:
\begin{equation}
H = -\frac{\hbar^2}{2M}\frac{d^2}{dx^2} + V_0 \sin^2\left(\frac{\pi x}{d_L}\right) + Fx \;.
\label{eq:1}
\end{equation}
 This Hamiltonian defines the well-known Wannier-Stark problem for the electrons moving within a crystal lattice in the presence of an external electric field \cite{Nenciu1991,Glueck2002,Holthaus2000}. For small Stark forces $F$, one can picture the evolution of a
momentum eigenstate induced by Eq.~(\ref{eq:1}) as an oscillatory motion in the ground energy band of the periodic lattice with Bloch period $T_{\rm B}$ \cite{Glueck2002,Holthaus2000,Morsch2006}, where 
\begin{equation}
T_{\rm B}=\frac{2\pi\hbar}{Fd_{\rm L}}.
\label{BlochPeriod}
\end{equation} 
\indent At stronger applied forces, a wave packet prepared in the ground band has a significant probability to tunnel at the band edge to the first excited band. This process of the quantum
tunnel across an energy gap at an avoided
crossing of the system's energy levels is described by the
Landau-Zener tunneling~\cite{Landau1932,Zener1932}. For a single tunneling event, the Landau-Zener tunneling probability is~\cite{Holthaus2000} 
\begin{equation}
{\rm P_{LZ}} = e^{ -\frac{\pi^2}{8F_0}\left(\frac{\Delta}{E_{\rm rec}}\right)^2 },
\label{eq:2}
\end{equation}
where we introduced the  $F_0$ dimensionless force
\begin{equation}
F_0= Fd_{\rm L}/E_{\rm rec}.
\label{F_0}
\end{equation}
  In the presence of a sequence of Landau-Zener tunneling events the Landau-Zener rate $\Gamma_{\rm {LZ}}$ to the excited band is obtained by multiplying ${\rm P_{LZ}}$ with the Bloch frequency $\nu_{\rm B} = 1/T_{\rm B}$~\cite{Glueck2002}. By introducing  the recoil frequency  $\nu_{\rm rec} = E_{\rm rec}/h$, $\Gamma_{\rm {LZ}}$ may be written
\begin{equation}
\Gamma_{\rm {LZ}} = \nu_{\rm rec} F_0 e^{ -\frac{\pi^2}{8F_0}\left(\frac{\Delta}{E_{\rm rec}}\right)^2 }\;. \label{eq:3}
\end{equation}
\indent For the  optical lattice periodic potential an alternative single-particle basis useful for describing the tunneling of particles among discrete lattice sites is provided by Wannier functions
~\cite{Ashcroft1976,Nenciu1991,Glueck2002,Holthaus2000,Bloch2008}. The $j$-th Wannier function $\ket{j}$ is centered around the $j$ lattice site, and the functions are orthonormal.  
In a given energy band the Hamiltonian for free motion on the periodic lattice is determined by hopping matrix elements, which in general connect lattice sites arbitrarily spaced.
However, because the hopping amplitude decreases rapidly with the distance, the tunneling Hamiltonian
may include only the  $J$ tunneling hopping between neighboring lattice sites 
\begin{equation}
H = \sum_{j}E_{\rm j}|j><j| - J \sum_{j} \left(|j><j+1|+|j+1><j|\right),
  \label{WannierStark}
\end{equation} 
where $E_{\rm j}$ defines the energy of the $j$-th site. 
For ultracold atoms  in an optical lattice with depth $V_0\gg E_{\rm rec}$, the nearest-neighbor tunneling energy $J$ is given by \cite{Zwerger2003}
\begin{equation}
J = \frac{4}{\sqrt{\pi}}E_{\rm rec}\left(\frac{V_0}{E_{\rm rec}}\right)^{3/4} \exp\left(-2\sqrt{\frac{V_0}{E_{\rm rec}}}\right) \,.
\label{tunneling}
\end{equation}
In the presence of an applied force $F$ supposing $E_j \equiv E_0=0$ the Hamiltonian becomes
\begin{equation}
H =Fd_L \sum_{j} j |j><j| - J \sum_{j} \left(|j><j+1|+|j+1><j|\right),
  \label{WannierStark_Force}
\end{equation}
However this Hamiltonian may be used to describe the atomic evolution in the ground band only when the Landau-Zener tunneling to the excited band can be neglected.  Fig.~\ref{LandauZenerLimit}  reports for a given value of the dimensionless force $F_0$, the $V_0$ optical depth where the hopping constant $J$ is ten times larger than $\Gamma_{\rm {LZ}}$.\\ 
\indent The simulation of the temporal evolution of the  Bose-Einstein condensate wavefunction is based either on the Gross-Pitaevskii equation based on a global mean-field description or on a many-body approach where the atomic number of the lattices sites is quantized~\cite{Pethick2002,Pitaevskii2003,Bloch2008}.  Apart from the theoretical results reported in Section V, we will concentrate here on the mean-field approach applied to describe experimental configurations and results reviewed in detail in Section IV. For a realistic description of those experiments, the Gross-Pitaevskii equation was used to simulate the temporal evolution of the condensate wave function $\psi(\vec{r},t)$ subjected to the optical lattice and to a confining harmonic potential, for instance with cylindrical symmetry
\begin{eqnarray}
\hspace{-4.5cm} i \hbar \frac{\partial }{\partial t}\psi (\vec{r},t)= \nonumber 
\\ \hspace{-1.5cm}
\left[-\frac{\hbar^2}{2M}\nabla ^2 + \frac{1}{2}M \left( \omega_x^2 x^2 +
\omega_{\rm r}^2 \rho^2 \right) + V_0 \sin^2\left(\frac{\pi x}{d_L}\right)
+ F x + gN \left| \psi(\vec{r},t) \right|^2
\right] \psi(\vec{r},t).
\label{eq:4}
\end{eqnarray}
The frequencies $\omega_x$ and $\omega_{\rm r}$ characterize the
longitudinal and transverse harmonic confinement. 
The atom-atom interactions
are modeled by the nonlinear term in Eq.~(\ref{eq:4}), with the
nonlinear coupling constant  given by $g=4\pi \hbar^2 a_s/M$,
where $a_s$ is the $s$-wave scattering length \cite{Pethick2002,Pitaevskii2003}.
Refs.~\cite{Cristiani2002,Morsch2006} introduced the $\tilde{g}$ dimensionless nonlinearity parameter
\begin{equation}
\tilde{g}=\frac{gn_0}{8E_{\rm rec}},
\label{nonlinearparameter}
\end{equation} 
computed from the peak density $n_0$ of the condensate initial state, to describe the nonlinear
coupling  relevant for optical lattice experiments. In the Thomas-Fermi regime
of the condensate~\cite{Pethick2002,Pitaevskii2003}, for given $\omega_x$ and
$\omega_r$ the density $n_0$, and therefore $\tilde{g}$, is proportional to
$N^{2/5}$ where $N$ is the number of atoms in the condensate. \\

\section{\label{sec3} Resonant tunneling in closed systems} 
\subsection{\label{sec3_1} Two levels}
\begin{figure}
\centering
\includegraphics[height=12cm]{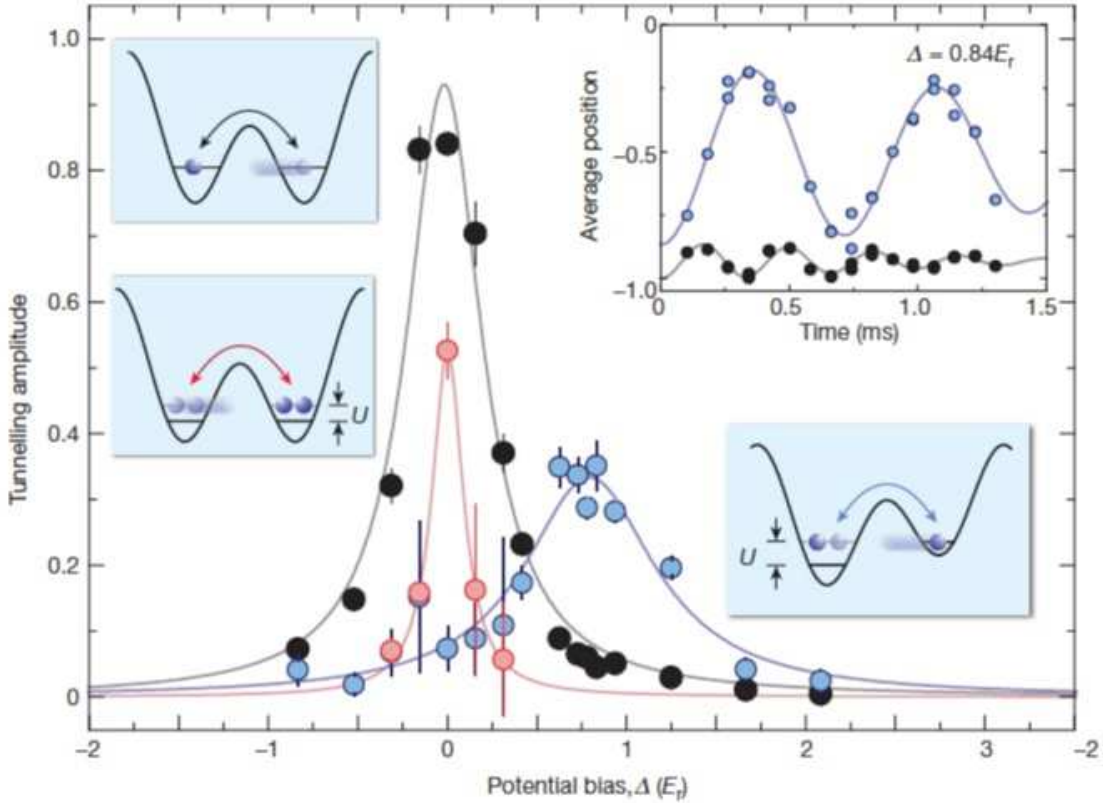}
\caption{Tunneling configuration and experimental results for the resonant tunneling of single and double atoms in a superlattice. The periodic double-well potential for ultracold rubidium atoms was realized by superimposing two periodic potentials with periodicities 
of $\lambda_{\rm L}=765.0$ nm (long lattice) and $\lambda_{\rm L}/2=382.5$ nm (short lattice), and controllable
intensities and relative phase. The depth was $V_0=12E_{\rm rec}$ for the short optical lattice, $V_0=9.5E_{\rm rec}$ for the long lattice. The upper left, lower left and lower right  insets describe the resonant tunneling configurations for one or two atoms per well. The upper left inset describes the oscillating motion of the atoms between the the two wells for the conditional resonant tunneling resonance where a single atom can tunnel only in the presence of a second atom and the interaction energy $U$ is matched by an applied bias. In the central part the amplitude of the tunneling Rabi oscillations, and the Lorentzian fit,  are shown as a function of the bias energy $\Delta$ for each of the tunneling configurations represented in the insets, black dots and Lorentzian centered at $\Delta=0$ for upper left one,  red dots and Lorentzian centered at $\Delta=0$ for lower left one, and blue dots and Lorentzian centered at $\Delta=0.78E_{\rm rec}$ for lower right one.  (From F\"olling {\em et al.}, Nature \cite{Foelling2007}. With permission by MacMillan).} 
\label{AsymmetricWell}
\end{figure}

Quantum tunneling of a two-level system takes place in the  double well potential. The quantum mechanical solution  shows that the wave packet initially localized in one of the wells performs oscillations between the two classically allowed region. The period of these oscillations is related to the inverse of the energy difference between the symmetric and antisymmetric quantum states of the double-well system, i.e. to the energy corresponding to the tunneling splitting. That energy is equal to the interaction Hamiltonian between the eigenstates of the two well.  In an asymmetric double well as  that shown in Fig.~\ref{DoubleBarrier}(a),  an energy matching between a ground state on one side and an excited state on the other side leads to a RET between those states.  The dashed lines in Fig \ref{AsymmetricWell}(a) denote the eigenenergies for the symmetric and antisymmetric quantum superposition of the wavefunctions in left and right wells. The tunneling evolution is described by the following Hamiltonian:
\begin{equation}
H = \sum_{j=1,2}E_{\rm j}|j><j| - J\left(|1><2|+|1><2|\right) +U \sum_{j=1,2}n_{\rm j}(n_{\rm j}-1),
  \label{doublewell}
\end{equation}
where $|1>$ and $|2>$ denote the wavefunctions of the resonant states in the left and right wells, $\Delta =E_1-E_2$ is the energy difference between the two wells, and $J$ is the tunneling energy, $U$ is the interatomic interaction energy and $n_{\rm j}$ is the atom number in the left or right well.   For the following analysis $U$ represents a shift in energy of the left or right well. By treating at first the  $U=0$  case, the atomic wavefunction may be expanded as a superposition of the $\ket{1,2}$ states 
\begin{equation}
|\Psi(t)> = \sum_{j=1,2}C_j(t)\ket{j},
  \label{wavefunction}
\end{equation}
the atomic evolution is characterized by Rabi oscillations between the two wells. For instance by supposing as initial condition $C_1(0)=1$ and  $C_2(0)=0$, the occupation probabilities of the left well at time $t$ are given by
\begin{eqnarray}
|C_2(t)|^2&=& \frac{J^2}{\Delta^2+J^2}\sin^2\sqrt{J^2+\frac{\Delta^2}{4}},\label {Cfinal}\\
|C_1(t)|^2&=&1-|C_2(t)|^2
  \label{Rabifrequency}
\end{eqnarray}
Therefore for the $\Delta=0$  resonance condition of RET, a complete  oscillation between the two wells at frequency $2J/\hbar$ takes place. The atomic interaction term $U$ shifting the $E_{\rm i=1,2}$ energies of the two wells may be included into the above equations for the occupation probabilities as a contribution to the $\Delta$ energy difference. Therefore the presence of the U interatomic energy modifies the RET condition.\\
\indent Periodic double-well structures may be created in properly chosen optical lattice or superlattice geometries. For cold atoms theoretical and experimental investigations were performed by a few authors~\cite{Castin1994,Dutta1999,Haycock2000,Teo2002}. For cold atoms the coherence length of the atomic wavefunction is comparable to the extent of each double-well, so that the long range periodicity of the optical lattice plays a minor role on the tunneling properties. Therefore those investigations will be mentioned here.  Those studies examined the new features appearing when the double-well potential depends on the internal atomic structure, for instance on the two electron spin states. This case was theoretically analyzed by Castin {\it et al.}~\cite{Castin1994}  within the context of two dimensional Sisyphus cooling.  Resonant tunneling between the adjacent potential wells of the periodic potential for the two internal states, not present in a 1D geometry, contribute with quantum processes to the cooling phenomena in optical lattices.  Dutta {\it et al.}~\cite{Dutta1999}  studied periodic well-to-well tunneling of $^{87}$Rb atoms on adiabatic potential surfaces of a 1D optical lattice. Atoms that tunnel between neighboring wells of
the lattice are an excellent tool for a careful study of topological potentials associated to the optical lattice. RET-like effects have been observed in a number of experiments to date. In Ref.~\cite{Teo2002}, resonant tunneling
was observed for cold atoms trapped by an optical lattice when an
applied magnetic field produced a Zeeman splitting of the energy
levels. At certain values of the applied magnetic field,  the
states in the up-shifting and down-shifting energy levels  were
tuned into resonance with one another. This led to RET drastically
altering the quantum dynamics of the system and producing a
modulation of the magnetization and lifetime of the atoms trapped
by the optical lattice.  Hacock {\it et al.}~\cite{Haycock2000} observed the quantum coherent dynamics of atomic spinor wave packets in the double-well potentials. With appropriate initial conditions the atomic system performed Rabi oscillations between the left and right localized states of the ground doublet, with the atomic wavepacket corresponding to a coherent superposition of these mesoscopically distinct quantum states. \\
\indent For ultracold atoms, Rabi oscillations in double well geometries have been investigated and  measured by the authors of~\cite{Foelling2007,Kierig2008}. A highly parallel structure of double wells is created  using optical lattice or optical superlattice configurations. In the superlattice configuration of \cite{Foelling2007} the periodic potentials created by two laser standing waves at wavelength $\lambda_{\rm L}$ and $\lambda_{\rm L}/2$ are applied to create a large set of individual wells. By changing the intensity of the standing wave lasers at the two wavelengths and their relative spatial phase, any configuration of symmetric or asymmetric double wells is created. In that experiment the double well investigation was performed with ultracold atoms in a Mott-insulator configuration having single atom occupation of the  wells~\cite{Bloch2008}. The modification of the optical lattice potential from a periodic structure of single wells to a periodic structure of double wells, by adiabatically raising an energy bump within each single well, 
allowed to produce the asymmetric loading of each double well.\\
\indent Fig.~\ref{AsymmetricWell} summarizes experimental results obtained in~\cite{Foelling2007} for the RET features in symmetric and asymmetric double wells. The tunneling of the ultracold atoms was measured as a function of the energy bias $\Delta$ between the wells.  The left upper inset schematizes the case of single atom tunneling.  The right lower one schematizes the tunneling of one atom in the presence of an energy shift produced by the atomic interaction ($U$ term in Eq.~(\ref{doublewell})). A conditional resonant tunneling resonance occurs, where a single atom can tunnel only in the presence of a second atom and the interaction energy $U$ is matched by the bias.  For these two cases  the measured atomic Rabi type dynamical evolution  between the two wells is shown the right upper inset. Because the presence of an atom in the left well shifts by $U$ the level energies, a bias $\Delta=-U$ is applied in order to compensate the shift. Thus, a resonant tunneling condition is verified and the blue data denote  the periodic occupation of the left well and right well, located at positions -1 and 0 respectively.  In the absence of an atom in the left well and without application of the bias,  the tunneling is not resonant and the Rabi oscillations take place with a reduced amplitude and at a higher frequency, in agreement with the description of Eqs.~(\ref{Cfinal}) and (\ref{Rabifrequency}). The left lower inset schematizes the case of a correlated atomic pair  tunneling, as produced in a second-order  tunneling process.The central part of that Figure reports the amplitude of the Rabi oscillations versus the $\Delta$ bias for the different tunneling configurations, and their fits by the Lorentzian line-shapes predicted by  Eq.~(\ref{Cfinal}).  The tunneling amplitude versus the potential bias is measured for
the case of single atoms (black data points) and initially doubly occupied lattice sites (blue and red data points). The blue data points and the Lorentzian fitted to the data point with center at  $\Delta=0.78(2) E_{\rm rec}$ correspond to the  conditional resonant tunneling resonance. The correlated pair tunneling (red circles) and the Lorentzian fit are resonant for zero bias because energies of both left and right wells are modified by the interaction energy $U$.\\ 
\indent While the previous description applies to single particle tunneling, quantum tunneling of macroscopic $N$-body atomic systems introduces qualitatively new aspects to the quantum evolution of  ultracold atoms, as investigated in \cite{DounasFrazer2007} for Bose-Einstein condensate in a tilted multilevel double-well potential. For a double-well without tilt as  experimentally investigated by Albiez {\it et al.}~\cite{Albiez2005}, the so-called self trapping regimes is realized where  the bosonic nonlinear interaction term of the Gross-Pitaevskii equation (\ref{eq:4}) modifies the level energies and inhibits the resonant tunneling between the wells.  Khomeriki {\it et al.}~\cite{Khomeriki2006} demonstrated for a double-well structure by a
pulse-wise change of the intermediate barrier height, it is possible to switch between the tunneling regime and the self-trapped one. 
\begin{figure}
\centering
\includegraphics[height=8cm]{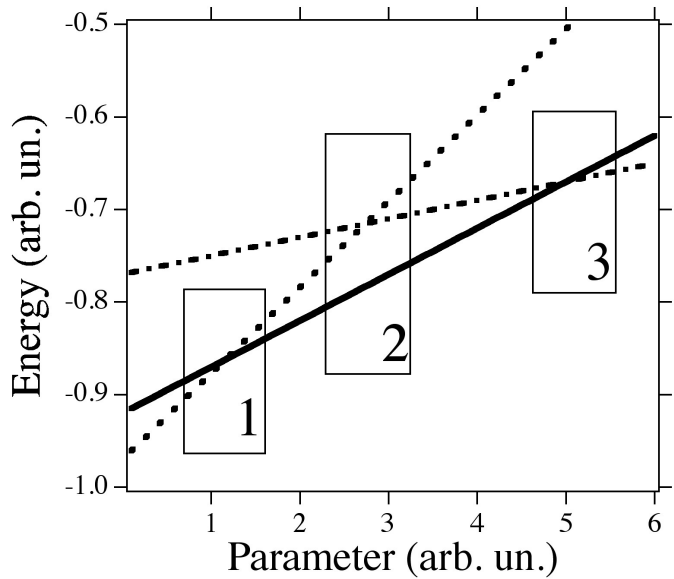}
\caption{Unperturbed energies $E_{\rm j}$, with $j=(1.3)$ (in arbitrary units) of three states experiencing crossings and anticrossings as a function of a parameter (also in arbitrary units).  Continuous lines corresponds to state $\ket{1}$, the dotted one to state $\ket{2}$ and the dot-dashed to state $\ket{3}$. The boxes marked with 1 and 3  denote regions where the tunneling is dominated by two-state interactions. The box marked 2 denotes a region where the three-state interaction may modify the tunneling rate between state $\ket{2}$ and $\ket{3}$. In region 3 without direct interaction between states $\ket{1}$ and $\ket{3}$ a locking of tunneling, corresponding to a level crossing with $E_1=E_3$, takes place.} 
\label{ThreeLevel}
\end{figure}
\subsection{\label{sec3_2} Three levels}
The idea of controlling the tunneling rate between two states has led several researchers
to consider the effect of external forces on the tunneling oscillations. Because the tunneling rate is related to the difference in the energies of the quantum states, a number of complicated scenarios arise when one of the states undergoes interaction with a third state, and that interaction may be controlled by an external parameter, for instance a magnetic or electric field. The tunneling wavepacket is described as a linear combination of the three initial states. Their interaction can drastically affect the eigenenergies of the Hamiltonian and it would be possible to explore different regimes, from strong suppression to enhancement of tunneling. \\
\indent This three-level control was theoretically investigated in Refs.~\cite{Averbukh2002,Hensinger2004} in connection to the dynamical tunneling produced by time dependent potentials  and for conditions as in an experiment by Raizen's group in 2001~\cite{Steck2001} and at NIST~\cite{Hensinger2001}. The tunneling period in the time-dependent systems is related to the differences between quasienergies of the Floquet states, just as the tunneling period in the time independent case has to do with the energy differences between the stationary states.  The experimental and theoretical investigations considered the case of the  tunneling doublet interacting with a third state associated with a chaotic region. The underlying classical phase space of the systems had a mixed regular-chaotic structure, giving the scenario of chaos-assisted \cite{Bohigas1993} or, more generally speaking, of dynamical tunneling \cite{DavisHeller1981}. \\
\indent We present here the basic of the three-level tunneling in the case of time independent potentials.  Fig.~\ref{ThreeLevel} schematizes the dependence on an external parameter for the  $E_{\rm j}$ energies for the $\ket{j}$ states, with $j=1\ldots3$, in the absence of interactions between them. We will discuss the modifications to those energies produced by atomic interactions between states,  supposing the presence of the interactions $U_{12}$ between states $\ket{1}$ and $\ket{2}$, and $U_{23}$ between states $\ket{2}$ and $\ket{3}$, and supposing no interaction between states $\ket{1}$ and $\ket{3}$. Notice that these interactions modify the  $E_{\rm j}$ energies in the regions close to the energy crossings, boxes 1, 2 and 3 in the Figure,  and that the tunneling frequency is determined by the splitting of the perturbed energies.  In the box with number 1 the $E_1-E_2$ energy separation, i.e. the tunneling, is dominated by the interaction between states $\ket{1}$ and $\ket{2}$. In the box denoted as 2, a three-state interaction takes place and the amplitude of the interaction between states $\ket{1}$ and $\ket{2}$ may be used to enhance or suppress the tunneling frequency between the states $\ket{2}$ and $\ket{3}$. Within the region denoted as 3,  in the absence of a direct interaction between the $\ket{1}$ and $\ket{3}$ states a $E_1=E_3$ crossing point exists. This crossing produces an absence of tunneling,  this configuration being indicated as locking of the wavefunction in the initial state of preparation~\cite{Averbukh2002}. 
\section{\label{sec4} Tunneling in open systems} 
\subsection{\label{sec4_1} Optical lattice without/with tilt}
\begin{figure}
\centering
\includegraphics[height=11.5cm]{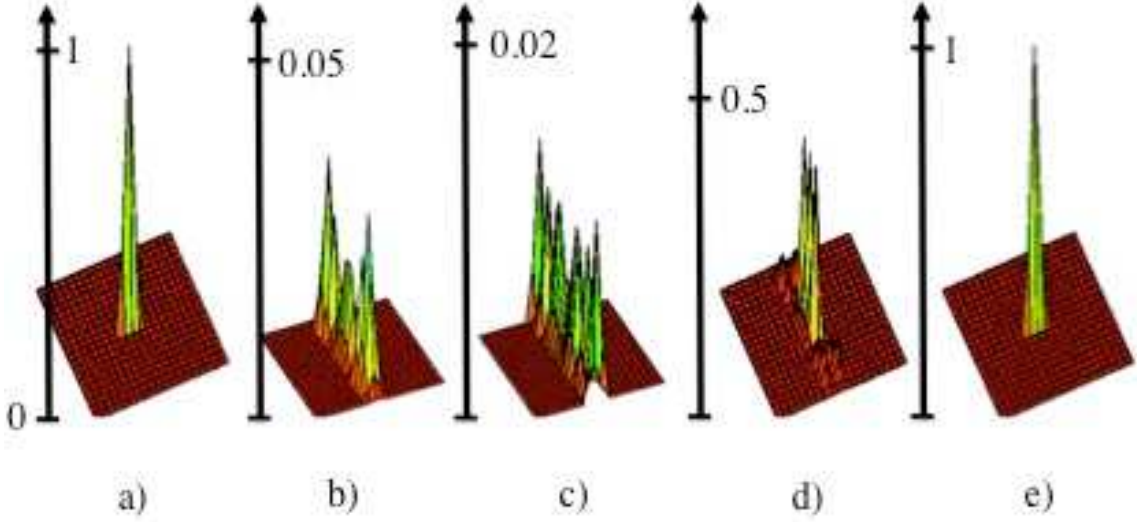}
\caption{Temporal recurrence of the occupation probability $|C_{\rm n}|^2$ versus the $n$ position of the lattice site at different interactions times, .  From (a) to (e) interaction  times are 0, 0.2, 0.5, 0.95, 1 measured in units of $T_{\rm R}$ The occupation probabilities are connected by lines.  Notice the reduced vertical scale at the intermediate times. The plots are obtained for the parameter $2JT_{\rm R}/\hbar=28$. }
\label{WaveRecurrence}
\end{figure}

An optical lattice is composed of an infinite number of neighboring wells uniformly distributed along one direction and spacing $d_L=\lambda/2$ between the minima, where $\lambda$ is the wavelength of the standing wave laser required to create the periodic potential for the atoms~\cite{Morsch2006}. This configuration corresponds to Fig.~\ref{TunnelingOL}(a). The  tunneling in this system has strong similarities to the double-well discussed above, when the presence of physical boundaries, as in the physical realizations, plays no role.  \\
\indent For a more general treatment we consider the case where an applied external force $F$ produces an energy difference $Fd_{\rm L}$ between neighboring wells, see Fig.~\ref{TunnelingOL}(b).  The  atomic evolution may be studied by considering  the localized Wannier wavefunction
$|i>$  and the perturbations originating from the atomic occupation  in neighboring sites \cite{Ashcroft1976}.  This approximation is valid when
the overlap of atomic wavefunctions introduces corrections to the localized atom picture, but  they are not  large enough to render the single site description irrelevant. The $H$ Wannier-Stark Hamiltonian determining the atomic evolution  in the absence of the interatomic interactions $U$ is given by
\begin{equation}
H = - J\sum_{j}\left(|j><j+1|+|j+1><j|\right)+Fd_L\sum_{j} j |j><j|.
  \label{WannierHamilt}
\end{equation}
In analogy to Eq.~(\ref{wavefunction}) the generic atomic wave function can be written as a superposition
of the $|j>$ localized wavefunctions where the sum extends over all  lattice sites.
The temporal evolution for the $C_{\rm i}$ coefficients  under the  Hamiltonian $H$ is given by
\begin{equation}
i\hbar\frac{dC_{\rm j}}{dt} =  jFd_LC_{\rm j}  - J\left(C_{\rm j+1} + C_{\rm j-1}\right),
\label{Ctemporal}
\end{equation}
and in the following the ground state energy $E_0$ will be supposed to be equal to zero. The solution of these coupled equations with $t=0$ initial condition of atomic occupation of the $i=0$ site, i.e. $C_{\rm j}(t=0)=\delta_{\rm j=0}$, leads to~\cite{Dunlap1986}  
\begin{equation}
|C_{\rm j} (t)|^2= {\cal J}_{\rm j}^2\left[\frac{2JT_{\rm R}}{\hbar} \sin \left(\frac{\pi t}{T_{\rm R}}\right)\right],
\label{occupation}
\end{equation}
having introduced the Bessel functions ${\cal J}_{\rm j}$ of $j-$th order.  The argument of the Bessel functions in Eq.~(\ref{occupation}) is an oscillatory  function of time. $T_{\rm R}$ represents the recurrence time for the evolution of the atomic wavefunction. For the present case of the resonant tunneling modified by the presence of a force $F$, $T_{\rm R}= T_{\rm B}$ whence the recurrence time coincides with the Bloch period $T_{\rm B}$ defined in Eq.~(\ref{BlochPeriod}) and is inversely proportional to the applied external force.  The temporal recurrence of the atomic wavefunction is  shown in Fig.~\ref{WaveRecurrence} for different times expressed in units of $T_{\rm R}$. Notice that the parameter $2JT_{\rm R}/\hbar$ of the Bessel function determines the range of lattice sites occupied by the periodic wavefunction expansion.   The corresponding atomic mean-square displacement is 
\begin{equation}
\frac{\sqrt{<m^2(t)>}}{d_{\rm L}}=\frac{\sqrt{2}JT_{\rm R}}{\pi \hbar}\left| \sin\left(\frac{\pi t}{T_{\rm R}}\right)\right|.
\label{displacement}
\end{equation}
In the limit of $Fd_{\rm L} \gg J$ the mean-square displacement is largely decreased because of the suppression of the resonant tunneling, as schematized in Fig.~\ref{TunnelingOL}(b). This suppression and the related Wannier-Stark localization of the wavefunction have been intensively discussed in the solid state physics theoretical literature ~\cite{Rossi1998,Glueck2002}. Korsch and coworkers~\cite{Korsch2003,Klumpp2007} have considered the case of an  atomic distribution not initially concentrated on a single site, and instead described by a Gaussian distribution with root mean-square $\sigma_0$. For that case  the temporal evolution of the mean-square displacement is given by
\begin{equation}
\frac{<m^2(t)>}{d^2_{\rm L}}=\left(\frac{\sigma_0}{d_{L}}\right)^2+2\left(\frac{JT_{\rm R}}{\pi \hbar}\right)^2 \sin^2\left(\frac{\pi t}{T_{\rm R}}\right)\left[1-e^{-\frac{d_L^2}{2\sigma_0^2}} \cos\left(\frac{2\pi t}{T_R}\right)-2e^{-\frac{d_L^2}{8\sigma_0^2}} \sin^2\left(\frac{\pi t}{T_R}\right)\right].
\label{displacementGaussian}
\end{equation}  
\indent In the absence of external force, taking the limit of $F\to0$, we recover the result of a diffusion process for the atomic wavefunction
\begin{eqnarray}
|C_{\rm j} (t)|^2&=& {\cal J}_{\rm j}^2\left[\frac{2Jt}{\hbar}\right], \label{occupationF0}\\
\frac{\sqrt{<m^2>}}{d_{\rm L}}&=&\frac{\sqrt{2}Jt}{\hbar}.
\label{meansquareF0}
\end{eqnarray}
\subsection{\label{sec4_n} Photon-assisted tunneling}
\begin{figure}
\centering
\includegraphics[height=8cm]{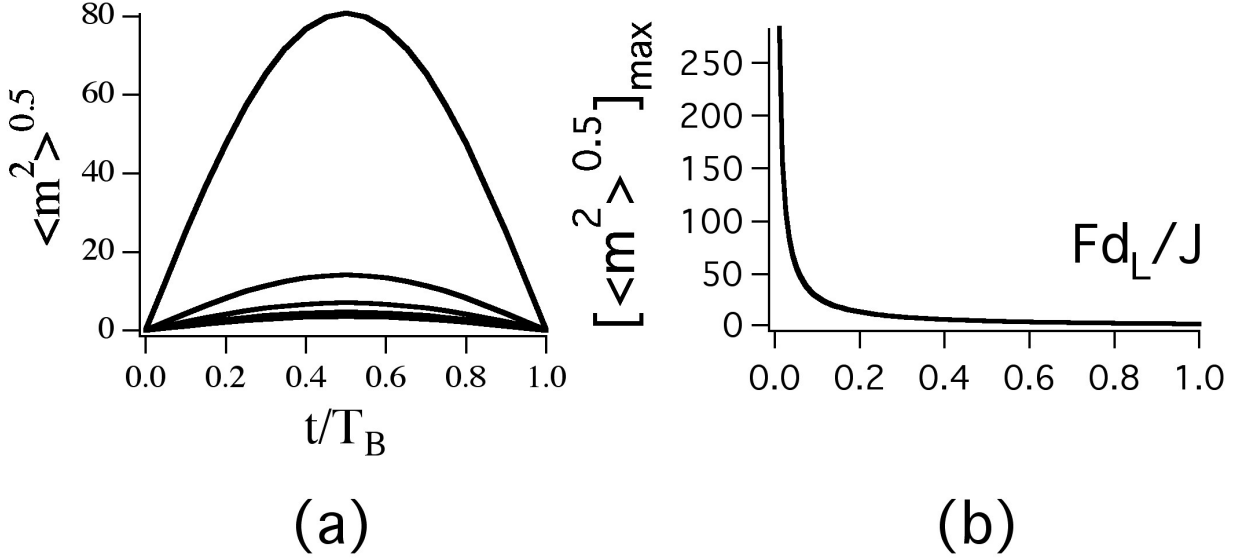}
\caption{(a) mean-square displacement versus time for different values of the unified RET energy mismatch, $Fd_{\rm L}/J$ for the Wannier-Stark localization and $\hbar \Delta \omega /J$ for the photon assisted tunneling. Results for values 0.2, 0.4, 0.6, 0.8 and 1.0 of the detuning parameter, with the displacement maximum decreasing at higher values. The time dependence of $\sqrt{<m^2>}$ is periodic in time with period  $T_{\rm B}$. In (b) the maximum of the mean-square displacement is plotted versus $Fd_{\rm L}/J$. The mean-square displacements are measured in units of the $d_{\rm L}$ lattice spacing.}
\label{WannierStarkdisplacement}
\end{figure}

The above analysis can be applied also to the photon-assisted tunneling occurring when the ground states of adjacent potential wells tuned out of resonance by the $Fd_{\rm L}$  static
potential are coupled by photons at frequency $\omega$ as schematized in Fig~\ref{TunnelingOL}(c). When the photon energy bridges the gap created by the static potential, tunneling is (partly) restored. The resonant tunneling is restored by a photon-assisted process when the energy provided by $n$ photons matches the separation energy $Fd_{\rm L}$ between neighboring wells. The energy resonance condition for the frequency $\omega_{\rm R}$ is given by 
\begin{equation}
n\hbar\omega_{\rm R}=Fd_L
\label{resonance}
\end{equation}
 with the integer $n$ denoting the order of the
photon-assisted resonance. This resonance  may be expressed as $\omega_{\rm R}=2\pi \nu_{\rm B}/n$ in terms of the Bloch frequency. The frequency detuning from  the resonance is $\Delta \omega = \omega -  \omega_{\rm R}$.  \\
\indent In solid state systems, the photons are typically in the microwave frequency range and the static potential is provided by an
electric bias field applied to the structure. Photon-assisted tunneling has been observed in superconducting diodes~\cite{Tien1963}, semiconductor superlattices~\cite{Keay1995,Keay1995b} and quantum dots~\cite{Kouwenhoven1994,Oosterkamp1997}.\\
\indent For the photon assisted tunneling of cold and ultracold atoms, a theoretical analysis was performed by Eckardt {\it et al.}~\cite{Eckardt2005} and by Kolovsky and Korsch~\cite{Kolovsky2010}, with experiments performed by Sias {\it et al.}~\cite{Sias2008}, Ivanov {\it et al.} ~\cite{Ivanov2008}, Alberti {\it et al.}~\cite{Alberti2009} and Haller {\it et al.}~\cite{Haller2010}.  In these experiments a periodic time-dependent potential was applied to the cold atoms through a periodic spatial oscillation of the optical lattice minima/maxima, to be referred to as shaking in the following.  In the lattice reference frame such a backward and forward motion of the periodic potential at frequency $\omega \approx \omega_{\rm R}$ along one direction is equivalent to a periodic force $F_{\omega}\cos(\omega t)$ applied to the atoms. Thus using the localized Wannier wavefunction introduced above for a deep lattice the atomic evolution is determined by the following Hamiltonian:
\begin{equation}
H_{\rm shaking} = -J\sum_{j}\left(|j><j+1|+|j+1><j|\right) +\left[Fd_L+ K \cos\left(\omega t\right)\right] \sum_{j} j |j><j|,
\label{shaking}
\end{equation}
once again not including  the $U$ interaction term. Here $K=F_\omega d_{\rm L}$, denoted as shaking amplitude, is the shaking energy difference between neighboring sites of the linear chain associated to the shaking. The theory of ref.~\cite{Eckardt2005} predicts that 
 when  the driving takes place at the frequency $\omega_{\rm R} \gg J/E_{\rm rec}$ and the resonance condition of Eq.~(\ref{resonance}) is satisfied, the shaking leads to an effective tunneling rate
\begin{equation}\label{eq2}
J_\mathrm{eff}(K,\omega_{\rm R})=J\mathcal{J}_{n}(\frac{K}{\hbar \omega_{\rm R}}).
\end{equation}
 Therefore a modification of the tunneling rate is obtained when the ratio of the rescaled shaking amplitude $K=F_\omega d_{\rm L}$ and 
the shaking frequency times $\hbar$ is varied. In the experimental realization \cite{Sias2008} the shaking frequency was fixed and the shaking amplitude was scanned to verify the relation of Eq.~(\ref{eq2}).\\
\indent   The previous analysis for the evolution of the atomic wavefunction under resonant tunneling can be applied also to the photon-assisted tunneling by using the approximation of a resonant dynamics introduced by Thommen {\it et al.}~\cite{Thommen2002} or equivalently by restricting our attention to the resonant Floquet states~\cite{Eckardt2008}.  In the presence of a driving at frequency $\omega$ and taking into account the static energy difference $Fd_{\rm L}$ between neighboring wells,  we write  for the atomic wavefunction  
\begin{equation}
|\Psi(t)> = \sum_{j,m}\tilde{C}_{\rm j,m}e^{-i\left(jFd_L+m\hbar\omega\right)t/\hbar}|j>,
  \label{wavefunctionFloquet}
\end{equation}
where the $j$ index labels the well and the $m$ index  the component in the Floquet spectrum.  For $\omega$ close to the $n$-th order resonance condition we may restrict the terms to the resonant ones in two sums of the above expansion
\begin{equation}
|\Psi(t)> = \sum_{j}e^{-ij\Delta \omega t}\tilde{C}^{\rm n}_{\rm j}|j>,
  \label{RedwavefunctionFloquet}
\end{equation}
where we have simplified the notation introducing the resonant coefficients  $\tilde{C}^{\rm n}_{\rm j}$. \\
\indent The temporal evolution of the  $\tilde{C}^{\rm n}_{\rm j}$ is described by an equation similar to Eq.~(\ref{Ctemporal}) where $J_\mathrm{eff}$ determines the tunneling energy of the $n$-th order resonance. Therefore, for the photon-assisted tunneling, the occupation of the $j$-th lattice site and the  mean square displacement of the atoms  are the analogues to those derived previously
\begin{eqnarray}
|C^{\rm n}_{\rm j} (t)|^2&=& {\cal J}_{\rm j}^2\left[\frac{2J_{\rm eff}T_{\rm R}}{\hbar}\sin \left(\frac{\pi t}{T_{\rm R}}\right)\right], \label{occupationPAT}\\
\frac{\sqrt{<m^2>}}{d_{\rm L}}=\frac{\sqrt{2}J_{\rm eff}T_{\rm R}}{\pi \hbar}\left| \sin \left(\frac{\pi t}{T_{\rm R}}\right)\right|,
\label{displacementPAT}
\end{eqnarray}  
with  $T_{\rm R}$ the recurrence time for this process  given by
\begin{equation}
T_{\rm R}= 2\pi/\Delta \omega.
\end{equation}
 This recurrence process was named as super-Bloch oscillations in refs.~\cite{Kolovsky2010,Haller2010}. 
For the resonant case $\Delta \omega=0$, the mean-square displacement is given by Eq.~(\ref{meansquareF0}) and the occupation probabilities are given by Eq. (\ref{occupationF0}).  Notice that for both Wannier-Stark localization and photon-assisted tunneling, the mean-square displacement and the occupation probabilities  have the same functional  dependence if we introduce a unifying parameter for the detuning from the resonant tunneling. This parameter is $Fd_{\rm L}$ for the case of an applied external force and $\hbar \Delta \omega$ for the case of the photon-assisted tunneling.  Thus, the data of Fig.~\ref{WaveRecurrence} applies also to the  occupation probabilities in the photon assisted tunneling. \\
\indent Few experiments on optical lattices have verified or made use of the theoretical predictions of this Section. In the following the experiments will be characterized by the depth $V_0$ of the optical lattice expressed in units $E_{\rm rec}$, and the photon-assisted frequency detuning $\Delta \omega_0$. \\
\indent The linear time dependence of atomic mean-square displacement predicted  by Eq.~(\ref{meansquareF0}) in the conditions of $F=0$ was applied by Lignier {\it et al.}~\cite{Lignier2007} to measure the $J$ tunneling energy and to verify that the experimental procedure reproduced the $J$ dependence on the lattice depth $V_0$ predicted by Eq.~(\ref{tunneling}).   The photon-assisted tunneling experiments~\cite{Sias2008,Ivanov2008}  made use of that linear dependence to measure the effective tunneling rate. In these experiments the linear dependence was tested for a total time larger than ten thousand tunneling times.   Notice that in all these experimental observations the initial distribution of the atomic wavefunction was not concentrated on a single well as in our theoretical analysis and instead covered several wells. Nevertheless a Gaussian convolution of the initial wavefunction spread and of the linearly  expanding mean-square displacement represented a good fit of the experimental observations, even at earlier times where the initial width is comparable to the tunneling spread. \\
\indent The Wannier-Stark localization of the atomic cloud in the presence of an applied force $F$ was examined by Sias {\it et al.}~\cite{Sias2008} as a reduction of the mean-square displacement  increasing the force amplitude at a given interrogation time.  Fig.~\ref{WannierStarkdisplacement}(a)  reports the temporal dependence of $\sqrt{<m^2>}/d_{\rm L}$ as predicted by Eq.~(\ref{displacement}), at different values of the  parameter $Fd_{\rm L}/J$ scanned in that experiment  within the interval (0,1). In order to provide a unified description the time is measured in units of $T_{\rm B}$. It appears that $\sqrt{<m^2>}$ is periodic in time with period $T_{\rm B}$ while the amplitude of the oscillation decreases with the force until the Wannier-Stark localization regime is reached where the atomic motion is blocked.  Fig.~\ref{WannierStarkdisplacement}(b)  shows the amplitude of the oscillation predicted by Eq.~(\ref{meansquareF0})versus the $Fd_{\rm L}/J$ parameter. By comparing this dependence to the Lorentzian one occurring for a two-level system of the previous Section, it appears that for an infinite systems of wells the oscillation amplitude decreases more rapidly increasing $Fd_{\rm L}/J$. For different values of the applied force, the maximum of the oscillation occurs at a different value of $t$. Therefore the  experiment of~\cite{Sias2008} that measured the oscillation amplitude  at a given interaction time, obtained results similar to those of Fig.~\ref{WannierStarkdisplacement}(b), not precisely fitted by the inverse law as sketched in Fig.~\ref{WannierStarkdisplacement}.\\
\indent For the photon-assisted tunneling  the  functional dependence on time  of the wavefunction spreading on the lattice and the mean-square displacement   was measured in~\cite{Alberti2009} for a total time equivalent up to seven recurrence times in the case of a drive detuned by $\Delta \omega/2\pi= \pm 5$ Hz and up to one recurrence time for the  $\Delta \omega/2\pi= \pm 0.260$ Hz detuning.  The measured sinusoidal evolutions are in reasonable agreement with the sinusoidal function predicted by our model and represented in Figs.~\ref{WaveRecurrence} and ~\ref{WannierStarkdisplacement}(a). Our model does not take into account the initial atomic distribution over several optical lattice sites, and in~\cite{Alberti2009}, because the atomic de Broglie wavelength was shorter than the lattice period,  the coherence degree among adjacent Wannier-Stark eigenstates was negligible. The quantum-mechanical evolution of the atomic wavefunction  under the tunneling Hamiltonian described by our analysis is limited by the presence of decoherence processes, and in \cite{Alberti2009}   a decoherence time of 28 seconds was measured. It would be interesting to investigate theoretically the  role of a decoherence process on the tunneling  evolution.  \\
\indent For the photon-assisted tunneling the mean-square amplitude dependence on the detuning $\Delta \omega$ is given by Eq.~(\ref{displacement}) with $T_{\rm R}=2 \pi/ \Delta \omega$. That functional dependence predicts that the fullwidth of the resonance line-shape  $\Delta \omega_{\rm FW}$, defined by the first zeros of the $\sin$ function, is determined by the experimental interrogation time $T$  
\begin{equation}
\Delta \omega_{\rm FW}=\frac{\pi}{T}.
\end{equation}
For interrogation times between 0.5 and 2 seconds of the experimental investigations line widths in the few Hertz range were measured. In the investigation of ~\cite{Ivanov2008} where the external force was gravity, the measurement of the resonance frequency  for the photon-assisted tunneling with the accuracy reached by the above interrogation time allowed those authors to measure the gravity acceleration with ppm resolution. This shows that sensitive RET effects have a great potential for applications, e.g. for precision measurements. \\
\indent The recurrence process of super-Bloch  oscillations was recently investigated by Haller {\it et al.}~\cite{Haller2010} for  $V_0/E_{\rm rec}$ values in the 3-7 range, and    $\Delta \omega/2\pi$ in the 0.1=2 Hz range. The recurrence oscillations were measured up to 2.5 seconds.

\begin{figure}
\centering
\includegraphics[height=6.5cm]{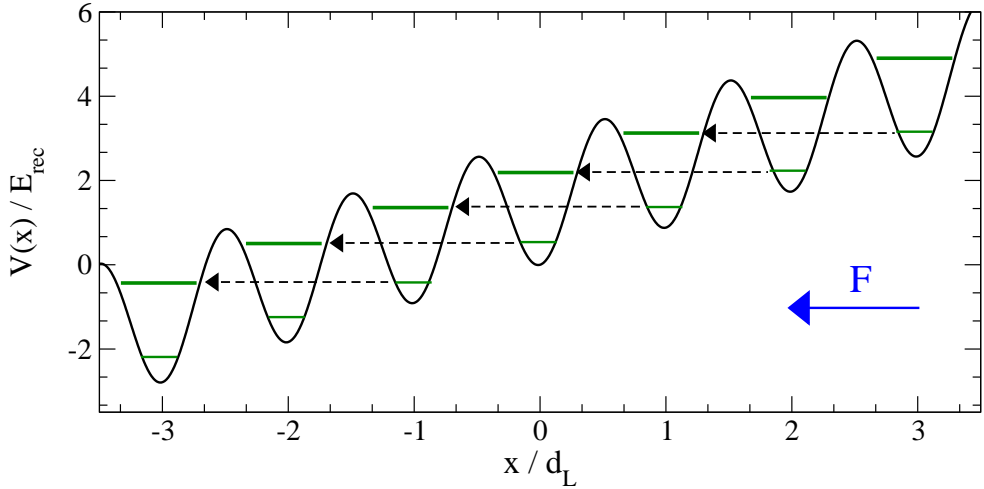}
\caption{Schematic of the RET process between second nearest
neighboring wells, i.e. for $\Delta i=2$. The tunneling of atoms
is resonantly enhanced when the energy difference between lattice
wells matches  the separation between the energy levels in
different potential wells. } \label{SchematicRET}
\end{figure}
\subsection{\label{sec4_2} RET in optical lattices with tilt}
 In spite of the fundamental RET nature and of its practical
interest, for a long time the experimental observation was restricted to
the motion of electrons in  superlattice structures \cite{Leo2003}. In 2007 Sias {\it et al.}~\cite{Sias2007} 
observed resonant tunneling using Bose-Einstein condensates in accelerated optical
lattice potentials. The nearly perfect control
over the parameters of this system allowed the authors to prepare the
condensates with arbitrary initial conditions and also to study
the effects of nonlinearity and a loss of coherence. Such observation 
can be generalized to studying noise and thermal effects in
resonant tunneling and underlines the usefulness of Bose-Einstein
condensates in optical lattices as model systems for the solid
state.\\
\indent A schematic representation of resonantly enhanced tunneling is
shown in Fig. \ref{SchematicRET}. In a tilted periodic potential, atoms can escape
by tunneling to the continuum via higher-lying levels. The tilt of
the potential is proportional to the applied force $F$ acting on the
atoms, and the tunneling rate $\Gamma_{LZ}$ can be calculated
using the Landau-Zener formula of Eq.~(\ref{eq:3}). The actual rates can dramatically 
deviate from Eq.~(\ref{eq:3}) when two Wannier-Stark levels in different potentials
wells are strongly coupled owing to the accidental degeneracy of Fig.~\ref{SchematicRET} where the
tilt-induced energy difference between wells $i$
and $i+\Delta i$ matches the separation between two quantized
energy levels, as pointed out for cold atoms by ~\cite{Bharucha1997}. Indeed, the tunneling probability can be
enhanced by a large factor over the Landau-Zener prediction (see theoretical and experimental results of Fig. \ref{lin}). \\
\indent By imposing an energy resonance between the Wannier-Stark levels in
different wells of an optical lattice shifted by the potential of
the external force, one finds that  the energy degeneracies  occur at
the values $F$ at which $Fd_L \Delta i$ ($\Delta i$ integer) is
close to the mean band gap between two coupled bands of the $F=0$
problem \cite{Glueck2002,Glutsch2004}. The actual peak positions
are slightly shifted with respect to this simplified estimate,
because the Wannier-Stark levels in the potential wells are only
approximately defined by the averaged band gap of the $F=0$
problem, a consequence of field-induced level shifts
\cite{Glueck2002}.

\subsubsection{ \label{sec4_4}Linear regime and decay rates}  
Although the finite and positive scattering length of $^{87}$Rb
atoms  means that the linear Hamiltonian of Eq.~(\ref{eq:1}) is never exactly realized in
experiments, the approximation of a non-interacting BEC is valid if 
the condensate density is maintained low. In that case, the interaction energy
can be made much smaller than all the other energy scales of the
system (recoil energy, band width, gap width) and hence it is negligible
for the present analysis of RET in a condensate. \\
\begin{figure}
\centering
\includegraphics[height=10cm]{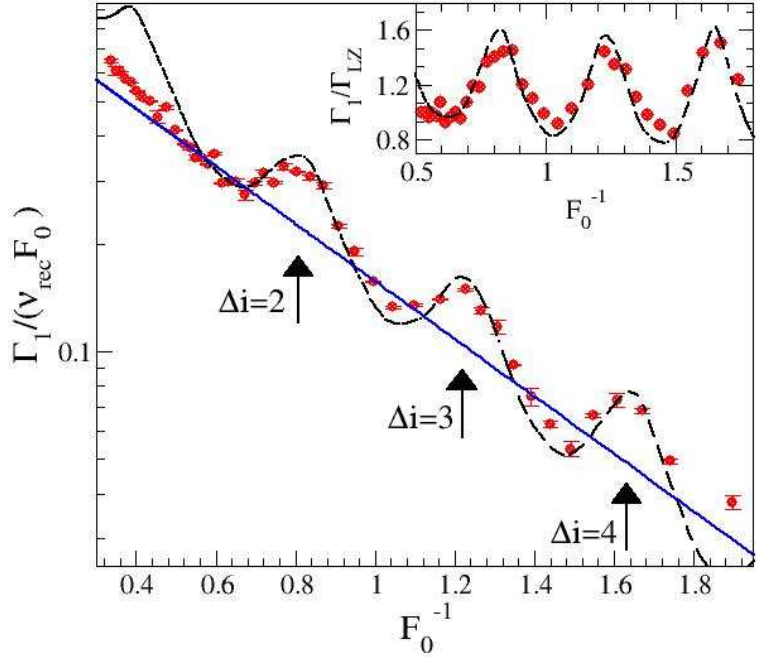}
\caption{Resonant tunneling in the linear regime. Shown here is
the tunneling rate from the lowest energy bands of the
lattice as a function of the normalized inverse force $F_0^{-1}$
for $V_0=2.5\,E_{rec}$ lattice depth. The straight line represents the prediction
of the Landau-Zener theory. Inset: Deviation from the Landau-Zener
prediction of Eq. (\ref{eq:2}). Adapted from Sias {\it et al.}~\cite{Sias2007}, Copyright 2007 
of American Physical Society.} 
\label{lin}
\end{figure}
\indent Fig.~\ref{lin} shows the results of~\cite{Sias2007} for experimental investigations with low-density
condensates and the nonlinearity parameter $\tilde{g}$ less than
$\approx 1\times 10^{-2}$, defined as the
limit of the linear regime. The tunneling rate
$\Gamma_{1}$ out of the $1$-th band  is shown as a function of
$F_0^{-1}$. Superimposed on the overall exponential dependence of
$\Gamma_1/F_0$ on $F_0^{-1}$, one clearly sees the resonant
tunneling peaks corresponding to the various resonances $\Delta
i=1,2,3,4$. Which of the resonances were visible in the
experiment depended on the choice of lattice parameters and the
finite experimental resolution. The limit $n=3$ for the highest
band explored in~\cite{Sias2007} was given by the maximum lattice depth
achievable. \\
\indent By measuring the positions of the $ \Delta i = 1,2,3$
tunneling resonances for different values of the lattice depth
$V_0$, it appeared that the resonances were
shifted according to the variation of the energy levels.
For deep enough lattices, the resonance positions may be derived from 
a numerical simulation but can also be
approximately calculated by making a harmonic approximation in the
lattice wells, which predicts a separation of the two lowest
energy levels ($n=1$ and $n=2$) of
\begin{equation}
\Delta E_{2-1} = 2E_\mathrm{rec}\sqrt{\frac{V_0}{E_\mathrm{rec}}}.
\end{equation}
By  imposing the resonance condition $\Delta E_{2-1}=F^{\rm res} d_L \Delta i$,
the  calculated $F^{\rm res}$ resonance position results in good approximation with that predicted in refs~\cite{Glueck2002,Wimberger2005}. \\
\begin{figure}
\centering
\includegraphics[height=8cm]{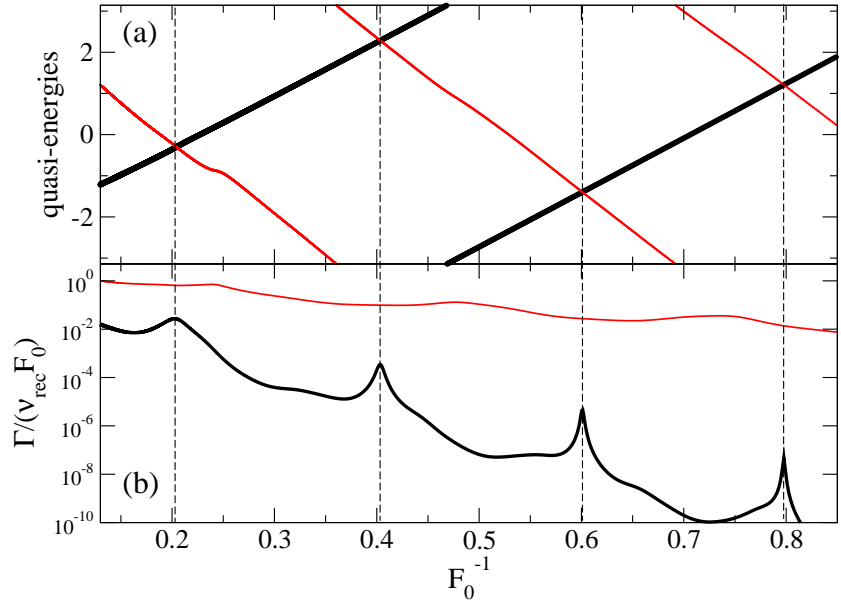}
\caption{In (a) real parts of the eigenenergies and in  (b) tunneling
rates for a lattice depth of $V_0/E_{\rm rec} = 10$ and the
Hamiltonian from Eq.~(\ref{eq:1}). The eigenenergies and the
tunneling rates are associated with two Wannier-Stark ladders or,
equivalently, with two energy bands: ground state (thick black
lines) and first excited state (thin red lines). The maxima of the
ground-state tunneling rates correspond to $\Delta i=1, 2, 3$, and $4$. Reproduced with permission from Zenesini {\it et al.} ~\cite{Zenesini2008}. Copyright Institute of Physics.}  \label{fig:1new}
\end{figure}\subsubsection{ \label{sec4_5}Avoided crossings} 
\indent The accessibility of higher energy levels allowed an experimental measurement of the tunneling 
rates around RET conditions of two strongly coupled bands.  The dependence of those rates on the system parameters was phrased into the frame of level crossing for states experiencing a loss rate. The modification of the level tunneling rate by the presence of a  degeneracy may be described  by a simple model of a two-level Hamiltonian with an energy separation $\epsilon$ described by an energy crossing splitting  $\epsilon=0$ and with a single level characterized by a decay rate \cite{Avron1982,Keck2003}. Real and imaginary parts of the Hamiltonian eigenvalues 
 are different for $\epsilon \ne 0$, and two different scenarios take place with crossings or anticrossings of the real and imaginary part of the Hamiltonian eigenvalues. In one case, denoted as type-I crossing, the imaginary parts of the
eigenvalues cross while the real parts anticross. In the second case, denoted as type-II crossing,
the eigenvalues anticross while the real parts cross. The numerical simulations of ref.~\cite{Zenesini2008} pointed out that the large majority of the RET explored experimentally correspond to type-II  crossings. As a consequence  if a resonance takes place between  the energy of the lower state and that of  the decaying upper level, the tunneling rate of the lower state
increases significantly. In addition the upper state experiences a
resonantly stabilized tunneling (RST) with a decrease of its tunneling rate. Fig.~\ref{fig:1new}(a) shows 
theoretical predictions for type-II crossing and anticrossings for the real parts of the eigenenergies  associated
with a RET configuration investigated experimentally as a function of
the experimental control parameter, the Stark force determined by the $F_0$ dimensionless parameter of Eq.~(\ref{F_0}).  The associated Wannier-Stark states tunneling 
rates are shown in Fig.~\ref{fig:1new}(b) as a function of  $F_0$. The strong
modulations on top of the global exponential decrease arise from
RET processes originated by the energy crossings. The resonance eigenstates and eigenenergies  for the non-interacting atoms described by Eq.~(\ref{eq:1})  were obtained in~\cite{Zenesini2008}  by diagonalizing an open version of the Hamiltonian \cite{Glueck1999,Glueck2002,Wimberger2006,Schlagheck2007,Witthaut2007}. \\
\begin{figure}[ht]
\includegraphics[width=13.5cm]{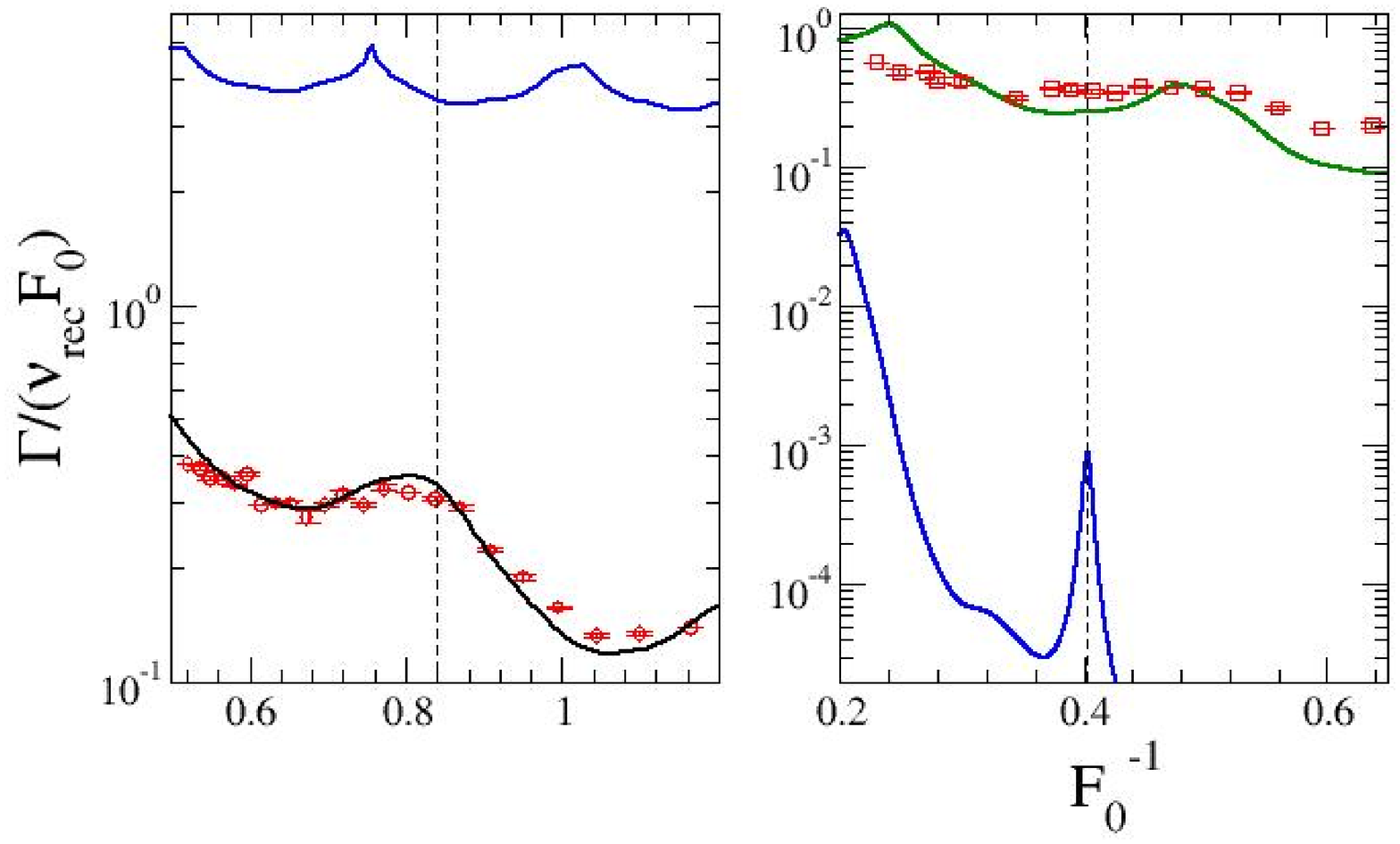}
\caption{\label{figure3} Anticrossing scenario of the RET rates.
(a) Theoretical plot of $\Gamma_{1,2}$ for
$V_0=2.5\,E_{\mathrm{rec}}$ with experimental points for
$\Gamma_1$. (b) Theoretical plot of $\Gamma_{1,2}$ for
$V_0=10\,E_{\mathrm{rec}}$ with experimental points for
$\Gamma_2$. Adapted from Sias {\it et al.}~\cite{Sias2007}.} \label{fig:ac}
\end{figure}
\indent Experimental data on anticrossings in the tunneling rates are in Fig.~\ref{fig:ac} taken from ref.~\cite{Sias2007}. Although a direct observation of the discussed anticrossing scenario  in two different levels
for the same set of parameters was not possible, the experimental investigation compared the ground and excited state tunneling rates $\Gamma_1$ and
$\Gamma_2$ with the theoretical predictions for two different parameter sets, as shown in Fig.~\ref{fig:ac}. This figure nicely reveals
the anticrossing of the corresponding tunneling rates of strongly coupled levels as a function of the control parameter $F_0$ around RET conditions.

\subsubsection{\label{sec4_3} Nonlinearity}
This Section discusses how the experimental investigation of RET in tilted optical lattices 
are modified by the atom-atom interactions in the Bose-Einstein condensate. 
We focus on a parameter regime where the Stark force essentially dominates the
dynamics of the condensate. Here the quantum tunneling between the
energy bands is significant and most easily detected
experimentally. The critical field values for which such
excitations are relevant can be estimated by comparing, for
instance, the potential energy difference between neighboring
wells, $F d_{\rm L}$, with the coupling parameters of the
many-body Bose-Hubbard model, i.e. the hopping constant $J$ and
interaction constant $U$ \cite{Morsch2006}.  \\
\indent Our analysis will exclude the regime
of  $F_0 \le J/E_{\rm rec} \approx U/E_{\rm rec}$ where a quantum chaotic system is
realized~\cite{Buchleitner2003,Thommen2003,Tomadin2007,Tomadin2008,Buonsante2008}.  The origin of 
quantum chaos, i.e. of the strongly force-dependent and
non-perturbative mixing of energy levels can be understood as a
consequence of the interaction-induced lifting of the degeneracy
of the multiparticle Wannier-Stark levels in the crossover regime
from Bloch to Wannier spectra, making nearby levels strongly
interact, for comparable magnitudes of hopping matrix elements and
Stark shifts. \\
\indent For the regime of $F_0 \gg J/E_{\rm rec}$, the effect
of weak atomic interactions is just a perturbative shifting and a small
splitting of many-body energy levels \cite{Wimberger2006,Tomadin2008}. In order to access the tunneling rates measured in the experiment of Sias {\it et al.}~\cite{Sias2007}, we determine
the temporal evolution of the survival probability $P_{\rm sur}(t)$ for the
condensate to remain in the energy band, in which it has been
prepared initially. As proposed in \cite{Wimberger2005} and applied 
in the experimental investigation, such a
survival probability is best measured in momentum space, since,
experimentally, the most easily measurable quantity is the
momentum distribution of the condensate obtained from a free
expansion after the evolution inside the lattice.
Such probability decays exponentially
\begin{equation}
P_{\rm sur}(t) = P_{\rm sur}(t=0) \exp \left(- \Gamma t \right).
\label{eqexp}
\end{equation} \\
\indent In the absence of interatomic interactions in the Gross-Pitaevskii equation, {\it e.g.} for nonlinearity 
parameter $g=0$ in Eq.~(\ref{eq:4}), the individual
tunneling events occurring when the condensate crosses the band
edge are independent.  Hence $P_{\rm sur}(t)$ globally, i.e. fitted over many Bloch periods, has a purely
exponential form, apart from the $t \to 0$ limit
\cite{Wilkinson1997}. When the nonlinear interaction term is
present, the condensate density decays with time too. As a consequence, the
rates $\Gamma$ are at best defined locally in time, and in the
presence of RET a sharp non-exponential decay may occur, as
discussed in \cite{Carr2005,Schlagheck2007}. Nevertheless, for short
evolution times and the weak nonlinear coupling strengths $\tilde{g}$ 
experimentally accessible ($\tilde{g}$ defined in Eq. (\ref{nonlinearparameter}), the global decay of the condensate 
is well fitted by an exponential law \cite{Wimberger2007,Sias2007}
\begin{equation}
P_{\rm sur}(t) = P_{\rm sur}(t=0) \exp \left(- \Gamma_n t \right),
\label{eq:surexp}
\end{equation}
with rates $\Gamma_n$ for the band $n=1$ (ground band), 2 (first excited band), 3 (second excited band), in which the atoms are initially prepared. \\ 
\indent We start our study of the tunneling rate in presence of a nonlinearity by discussing the position of RET peaks. These peaks, whose positions for the single-particle evolution are
studied in the previous part of this 
Section \ref{sec4_2}, are affected by the nonlinear
interaction term appearing in the Gross-Pitaevskii Eq.~(\ref{eq:4}) for BEC. The RET resonances originate from an
exact matching of energy levels in neighboring potential wells,
and hence they are very sensitive to slight perturbations. A shift of the RET peaks in
energy or in the position of the Stark force, predicted in
\cite{Wimberger2006} for large value of the $\tilde{g}$ parameter, is negligible for
the experimental investigated nonlinearities $\tilde{g} < 0.06$, the resonance  shift 
corresponding to the extremely small $\Delta F_0 < 5 \times
10^{-4}$ value~\cite{Wimberger2006}.\\
\begin{figure}
\centering
\includegraphics[height=14cm]{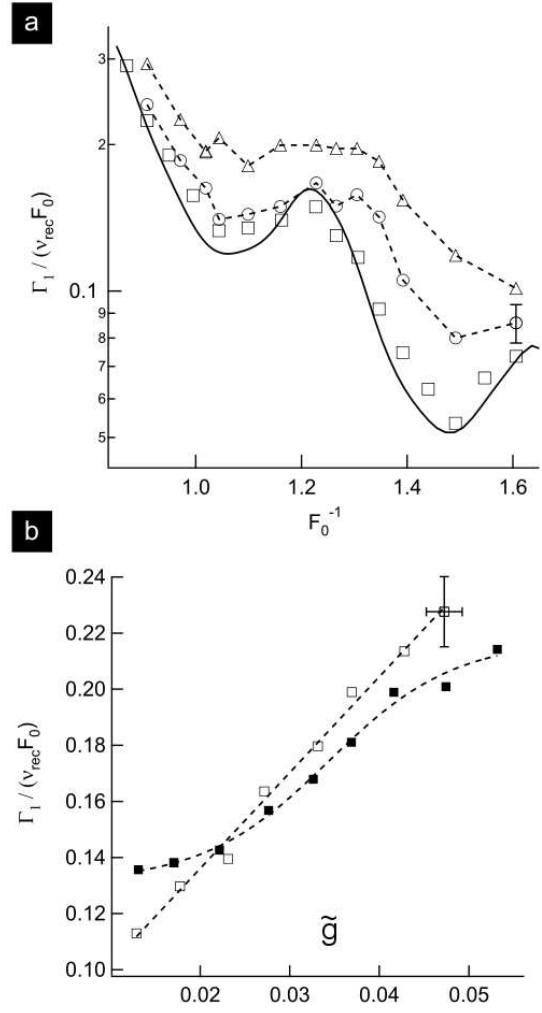}
\caption{Resonant tunneling in the nonlinear regime. (a) The
tunneling rates for $\Delta i = 2$ from the lowest energy band of the optical lattice
as a function of the normalized inverse force $F_0^{-1}$ for a
lattice depth  $V_0/E_\mathrm{rec}=3.5$  and different values of the nonlinearity parameter,
$\tilde{g}\approx 0.01, 0.022,0.033$ from bottom to top. The continuous line
is the theoretical prediction in the linear regime. The dashed lines connect the data obtained 
at large $\tilde{g}$ values. As the
nonlinearity increases, the overall tunneling rate increases and
the resonance peak becomes less pronounced. (b) Dependence of the tunneling rate 
on the nonlinear parameter $\tilde{g}$   at the position  $F_0^{-1} = 0.71$ (solid symbols)
of the RET spectrum peak and at $F_0^{-1} = 0.60$ (open symbols) a  the RET spectrum local minimum,
 for $V_0/E_\mathrm{rec}=3.0$. Adapted from Zenesini {\it et al.}~\cite{Zenesini2008}.}
\label{nonlin}.
\end{figure}
\indent The $\tilde{g}\gtrsim 1\times 10^{-2}$ regime  
was entered by carrying out the acceleration experiments in radially tighter
traps (radial frequency $\gtrsim 100\,\mathrm{Hz}$) and hence at
larger condensate densities. Fig.~\ref{nonlin}(a) shows the $\Delta i=2$ and
$\Delta i=3$ resonance peaks of the ground-state band ($n=1$) for
increasing values of $\tilde{g}$, starting from the linear case and going
up to $\tilde{g}\approx3\times 10^{-2}$. As the nonlinearity increases,
two effects occur. First, the overall (off-resonant) level of
$\Gamma_1$ increases linearly with $\tilde{g}$. This is in agreement with
earlier experiments on nonlinear Landau-Zener
tunneling~\cite{Morsch2001,JonaLasinio2003} and can be modeled by
a condensate evolution taking place within a nonlinearity-dependent effective potential
$V_{\mathrm{eff}}=V_0/(1+4\tilde{g})$~\cite{ChoiNiu1999}. Second, with
increasing nonlinearity, the contrast of the RET peak is decreased
and the peak eventually vanishes, as evident from the
different on-resonance and off-resonance dependence of the
tunneling rate as a function of the atom number $N$ (and hence the
nonlinearity), c.f. Fig.~\ref{nonlin} (b).  \\
\indent The critical value of $\tilde{g}$ for which
the nonlinearity affects the resonance peak is
estimated by comparing the width of the RET peaks of a band $n$ 
(which essentially is determined by the tunneling width  $\Gamma_{\rm n+1}$ of the band into 
which the atoms tunnel) with the energy scale of the
nonlinearity. In the experimental investigation of Sias {\it et al.}~\cite{Sias2007} atomic nonlinearities corresponding to
this order-of-magnitude argument were reached.   For the parameters of Fig.~\ref{lin}
and \ref{nonlin}(a) and the RET peak with $\Delta i = 2$, the typical width $\Gamma_2$ of the decaying state to 
which the atoms tunneling energy is of the order of $0.2 \ldots 0.5 \times E_\mathrm{rec}$. Since $\tilde{g}$ reflects the nonlinearity
expressed in units of $8\times E_\mathrm{rec}$, this means that substantial deviations from the linear behavior are expected
when $\tilde{g}\gtrsim 0.025 \ldots 0.06$. The experimental observations confirmed that this
threshold is a good estimate for the onset of the destruction of the 
RET peak, observed to occur around $\tilde{g}=0.02$ in Fig.~\ref{nonlin}(a).\\
\indent The role of nonlinearity on the time evolution of an Wannier-Stark state localized in a single site of the optical lattice was also studied 
by Krimer {\it et al.}~\cite{Krimer2009}. They predict that the nonlinearity strength leads to different regimes, where the nonlinearity induced shift in
 the energy of the lattice may enhance or inhibit RET.

\section{\label{sec5} Many-body tunneling}

\begin{figure} 
  \centering 
  \includegraphics[width=12cm]{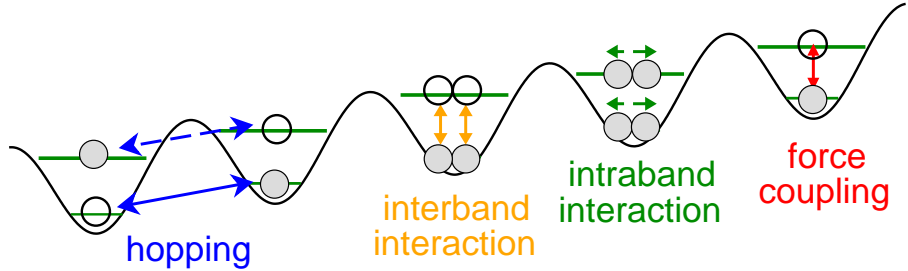}
  \caption{Sketch of most of the terms of the Hamiltonian (\ref{eq:two}). This model can be used to {\em fully} describe RET, since it contains
           excited levels in each potential well, in contrast to the effective model of Section \ref{sec5_1}.
}
  \label{fig:2bands} 
\end{figure} 

In state-of-the-art experiments the interatomic interactions can be tuned by the transversal confinement and by Feshbach resonances \cite{Bloch2008}, resulting in strong interaction-induced correlations. A good starting point for the discussion of true many-body effects is to use a lattice model, as introduced above for a single particle, c.f. Eq.~(\ref{WannierStark_Force}), and widely used in the context of strongly-correlated ultracold quantum gases \cite{Bloch2008}. Such a lattice description has the great advantage that the number of degrees of freedom automatically is bounded as compared to field theoretical approaches (see, e.g. \cite{Kuehner1998,Duine2003} and Refs. therein), and one can use it for practical numerical simulations. \\
\indent Using a single-band model, the regime of strong correlations in the Wannier-Stark system was addressed in \cite{Buchleitner2003,Kolovsky2003,Tomadin2007,Tomadin2008,Buonsante2008}, revealing the sensitive dependence of the system's dynamics on the Stark force $F$. The single-band Bose-Hubbard system of \cite{Buchleitner2003,Kolovsky2003} is defined by the following Hamiltonian with the creation $a_{l}^\dagger$, annihilation $a_{l}$, and number operators $n_{l}^a$ for the first band of a lattice with sites $l=1\ldots L$:
\begin{equation}
H_{\rm 1B} = \sum_{l=1}^{L} \left[ F_0 E_{\rm rec} l n_l^a - \frac{J_a}{2} \left ( a_{l+1}^\dagger a_{l} + \text{h.c.}  \right ) + \frac{U_a}{2} n_{l}^a (n_l^a - 1) 
+ \epsilon_a n_l^a\right], \label{eq:single} 
\end{equation}
where the last term describes the on site energy. \\
\indent In order to describe interband tunneling and phenomena related to those discussed in the previous Section IV, such a model has to be extended to include at least the equivalent of two single-particle energy bands (as plotted in Fig.~\ref{bands}). In the presence of strong interatomic interactions parameterized by $U$ terms, the single-band model of Eq. (\ref{eq:single}) should be extended to allow for interband transitions, as e.g. realized at $F_0=0$ in experiments with fermionic interacting atoms \cite{Koehl2005}. Doing so, the authors of \cite{Tomadin2008,Ploetz2010} arrived at the following full model Hamiltonian for a closed two-band system schematically sketched  in Fig.~\ref{fig:2bands}:
\begin{eqnarray}
       H(t)  ~ = ~ \epsilon_a \sum_{l=1}^{L} n_l^a + \epsilon_b \sum_{l=1}^{L} n_l^b  \quad \text{onsite energy} \nonumber \\ 
             ~ + ~ F_0 D E_{\rm rec} \sum_{l=1}^{L} (b_l^{\dagger}a_l + \text{h.c.}) \quad \text{force coupling} \nonumber \\
             ~ - ~ \frac{1}{2}J_a \sum_{l=1} (\text{e}^{i \frac{2 \pi t}{T_{\rm B}}}a_{l+1}^{\dagger}a_l + \text{h.c.}) ~ + ~ \frac{1}{2}J_b \sum_{l} (\text{e}^{i \frac{2 \pi t}{T_{\rm B}}}b_{l+1}^{\dagger}b_l + \text{h.c.}) \quad \text{hopping in the bands} \nonumber \\
         ~ + ~ \frac{1}{2}U_a \sum_{l=1}^{L} n_{l}^a (n_l^a-1) ~ + ~ \frac{1}{2}U_b \sum_{l=1} n_{l}^b (n_l^b - 1) \qquad\text{onsite interaction} \nonumber \\
                 ~ + ~ 2 U_x \sum_{l=1}^{L} n_{l}^a n_l^b ~ + ~ \frac{1}{2}U_x \sum_{l=1}^{L} (b_l^{\dagger}b_l^{\dagger}a_l a_l + \text{h.c.} )  \quad \text{interband interaction},
\label{eq:two}
\end{eqnarray}
where the $b$ index  and the $b_l,b_l^\dagger$ creation/annihilation operators are associated to the terms of the second band. $D$ is the "dipole" matrix element between the ground and excited single-particle states in a single lattice site (measured in $2\pi/d_{\rm L}$ length units, c.f. the appendix A of \cite{Tomadin2008} for a detailed explanation of how parameters are computed from the physical model).\\
\indent Within this full two-band system, {\it two} dominating mechanisms  promote  to the second band particles starting from the ground band. The first one is a single-particle coupling arising from the force term
\begin{equation}
H_{1} = F_0 D E_{\rm rec}\sum_{l=1}^{L}\left ( b_l^{\dagger}a_l + \text{h.c.}  \right ) \;,
\label{eq:perone}
\end{equation}
where the dipole matrix element $D$ depends only on the lattice depth $V_0$ (measured in recoil energies
according to the definition above, c.f., Eq.~(\ref{recoilenergy})). The second one is a many-body effect, 
describing cotunneling of two particles from  the first band into  the second band
\begin{equation}
H_{2} =\frac{U_x}{2} \sum_{l=1}^{L}\left (  b_l^{\dagger}b_l^{\dagger}a_l a_l + \text{h.c.}  \right ).
\label{eq:pertwo}
\end{equation}
In Eq.~(\ref{eq:two}) the tilting terms arising from the Stark force $F_0$ have been transformed into a phase factor $\text{e}^{\pm i 2\pi t/T_{\rm B}}$ for the hopping terms by changing into the accelerated frame of reference \cite{Kolovsky2003}. This transformation nicely shows that the present problem is intrinsically time-dependent. Since $H(t)=H(t+T_{\rm B})$  is periodic with the Bloch period $T_{\rm B}$, a Floquet analysis can be used to  derive the eigenbasis of the one-period evolution operator generated by $H(t)$. This trick allows also the application of periodic boundary conditions, which is reasonable in order to model large experimental systems, typically extending over a large number of lattice sites. The Hamiltonian of Eq. (\ref{eq:two}) contains hopping terms linking nearest neighboring wells in both bands ($J_a$ and $J_b$), and terms couplings different  bands at a  fixed lattice site $l$ either by the force presence ($F_0 D$) or by interactions ($U_x$). Other terms can, in principle, be included, yet they turn out to be exponentially suppressed for sufficiently deep lattices which are well described by Bose-Hubbard like models \cite{Bloch2008}. \\
\indent Because of its complex form and the large number of participating many-particle states, the above Hamiltonian is hard to interprete and to treat even numerically, for reasonable numbers of atoms $N$ and lattice sites $L$. Two approximate treatments will be presented in the following. Section \ref{sec5_1} uses an effective one band model which nevertheless takes the coupling terms between the bands of (\ref{eq:two}) into account. While this model is valid for small interband couplings, Section \ref{sec5_2} presents analytical and numerical results for the full model (\ref{eq:two}), which on the other hand is valid for arbitrary interband couplings but is perturbative in the atom-atom interaction terms $U_a, U_b$, and $U_x$.

\begin{figure} 
  \centering 
  \includegraphics[width=12.5cm]{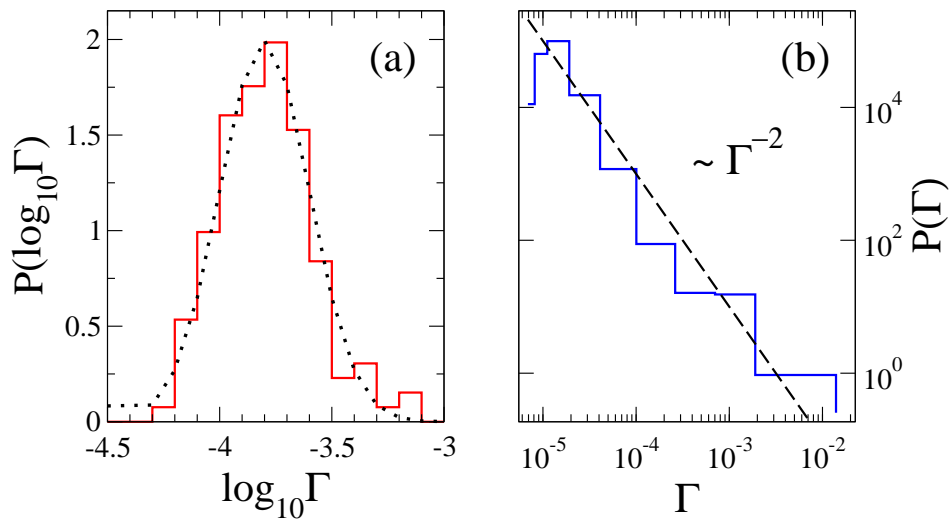}
  \caption{Rate distributions for the spectrum of an open one-band Bose-Hubbard model  in (a)  for 
            $F_0 \simeq 0.47, J_a/E_{\rm rec} =0.22, U_a/E_{\rm rec} = 0.2, U_x /E_{\rm rec} \simeq 0.1$ (for system size $(N,L)=(7,6)$)
           and in (b) for $F_0 \simeq 0.17, J_a/E_{\rm rec} =0.22, U_a/E_{\rm rec} = 0.2, U_x /E_{\rm rec} \simeq 0.1$ ($(N,L)=(9,8)$). 
           In the regime where the Stark force dominates 
           a log-normal distribution fits well the data (dotted in (a)), whilst a
           power-law P$(\Gamma) \propto \Gamma^{-x}$ distribution is found with $x \approx 2$  in the
           strongly coupled case (dashed line in (b)).
}
  \label{fig:gammas} 
\end{figure}

\subsection{\label{sec5_1} Open one-band model}
 Instead of using a numerically hardly tractable complete many-band model, we introduce here a perturbative decay of the many-particle modes in the ground band to a second energy band. This novel approach when applied to the Landau-Zener-like tunneling between the first and the second band \cite{Bharucha1997,Morsch2001,Cristiani2002,JonaLasinio2003,Wimberger2005,Sias2007} predicts the expected tunneling rates and their statistical distributions. \\
\indent To justify this perturbative approach, it is crucial to realize that the terms of Eqs. (\ref{eq:perone}) and (\ref{eq:pertwo}) must be small compared with the band gap $\Delta E \equiv \epsilon_b - \epsilon_a$ and indeed $F_0 D, U_x \ll \Delta E$ for the parameters of Fig.~\ref{fig:gammas}.
As exercised in detail by Tomadin {\em et al.}~\cite{Tomadin2007,Tomadin2008}, from these two coupling terms by using Fermi's golden rule one can compute analytically the corresponding tunneling rates $\Gamma_1(s)$ and $\Gamma_2(s)$ for each basis state labeled by $s$. Those  rates allow the computation of  the total width $\Gamma (s)=\Gamma_{1}(s)+\Gamma_{2}(s)$ defined by the two analyzed coupling processes for each basis state $\ket{s}$ of the single-band problem given in Eq.~(\ref{eq:single}). The $\Gamma (s)$ are inserted as complex potentials in the diagonal of the single-band Hamiltonian matrix. Along with the statistics of the level spacings defined by the real parts of its eigenspectrum $\mathsf{Re}\,\{E_{j}\}$ studied in \cite{Buchleitner2003,Kolovsky2003,Tomadin2007,Tomadin2008,Buonsante2008}, the statistical distributions of the tunneling rates $\Gamma_{j}=-2\mathsf{Im}\,\{E_{j}\}$ may be analyzed, as done in Fig.~\ref{fig:gammas}. For the regime where the motion of the atoms is localized along the lattice \cite{Glueck2002} that distribution is in good agreement with the expected log-normal distribution of tunneling rates (or of the similarly behaving conductance) \cite{Beenakker1997}. In that regime the Stark force dominates and the system shows nearly perfect single-particle Bloch oscillations \cite{Buchleitner2003}, the distributions agreing with those predicted from the localization theory \cite{Beenakker1997,Kottos2005}. On the other hand, 
when the Stark force is comparable with $J_a$ and $U_a$ and all modes of our Bose-Hubbard model are strongly coupled, the rate distribution of Fig.~\ref{fig:gammas}(b) follows the expected power-law for open quantum chaotic systems in the diffusive regime \cite{Kottos2005}.  This regime shows strong signatures of quantum chaos \cite{Buchleitner2003,Kolovsky2003,Tomadin2007,Tomadin2008,Buonsante2008}, which manifest  also in the rate distributions \cite{Tomadin2007,Tomadin2008}.

\subsection{\label{sec5_2} Closed two-band model}

\begin{figure} 
  \centering 
  \includegraphics[width=14.5cm]{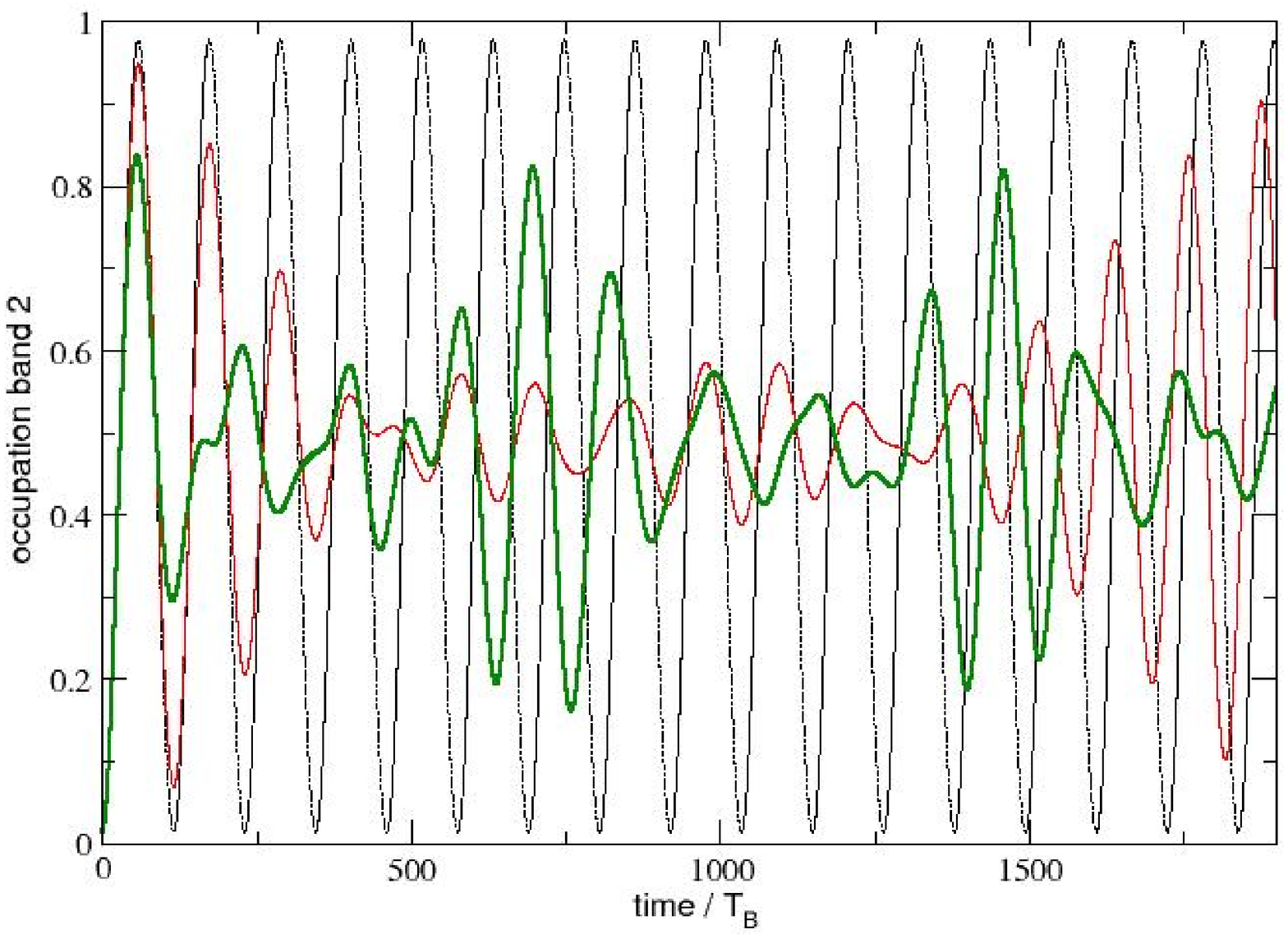}
  \caption{Population in the upper band as a function of time for the rescaling parameter $\alpha=0$ (black dotted line), 0.2 (faint red line) and 0.5 (green thick line) in a closed two-band model. Clearly visible are the interaction induced collapes and revivals of the RET oscillations between the bands. Other parameters are $F_0=1.87$ (dominating energy scale!) and
$J_a=0.1, J_b=0.77, U_a=0.023, U_b=0.014, U_x=0.01,  \epsilon_b-\epsilon_a=3.38$ (all in recoil energy units) and $D=-0.16$ in length units, and $(N,L)=(5,6)$.
}
  \label{fig:rabi} 
\end{figure}

\begin{figure} 
  \centering 
  \includegraphics[width=12cm]{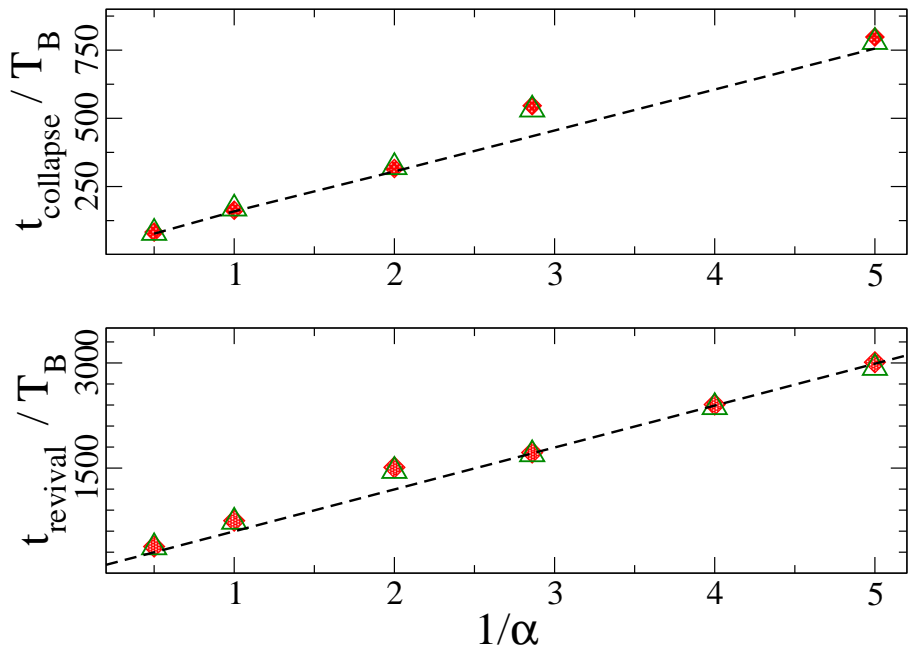}
  \caption{Collape and revival times extracted from data (symbols for two different system parameter sets) as shown in Fig.~\ref{fig:rabi} versus the inverse of the atom-atom interaction rescaling factor $\alpha$. As expected for a two-state Rabi problem perturbed by a coupling to additional states, both times scale inverse proportionally to $\alpha$. The dashed lines should guide the eye.
}
  \label{fig:scaling} 
\end{figure}

Since the model introduced in the previous Section \ref{sec5_1} cannot account for resonant tunneling between a ground level of one well and an excited level of another well, a different model which applies also for strong transitions between the bands was investigated by Pl\"otz {\it et al.}~\cite{Ploetz2010}. This model is based on the full Hamiltonian of Eq. (\ref{eq:two}) sketched schematically in Fig.~\ref{fig:2bands}. \\
\indent When the Stark force is tuned to the value where RET occurs for the single particle problem (c.f. Section \ref{sec4_2}), the strong coupling of the atoms prepared in the ground band into the excited band plays an important role. Since the model is closed, i.e. higher bands are neglected, there is no asymptotic tunneling as in the experimental situation described in Section \ref{sec4_2}. As a consequence, we observe an oscillation of the probability of occupying the lower and upper band, respectively, which is particularly pronounced at RET conditions. For a single particle in our lattice model, such RET oscillations can be understood easily, since in Floquet space (remembering that our Hamiltonian of Eq.~(\ref{eq:two}) is periodically time-dependent) the problem reduces to an effective two state model of resonantly coupled states \cite{Nakamura2001,Ploetz2010}. In this effective description, the evolution corresponds to the two level Rabi problem of quantum optics \cite{Meystre2007}. For non-vanishing atom-atom interaction, the situation complicates, of course, and we expect a degradation of those single-particle Rabi oscillations. This is illustrated in Fig.~\ref{fig:rabi}. 
The period of the single-particle interband oscillation is given by the following formula derived in \cite{Ploetz2010}:
\begin{equation}
\frac{ t_{\rm osc} } { T_{\rm Bloch} } \approx \frac{1}{\left| 2 D J_{\Delta i} \left(\frac{J_b-J_a}{F_0} \right)\right|}\,,
\label{eq:osc}
\end{equation}
where $\Delta i$ is the resonance order introduced in Section \ref{sec4_2} and $J_{\Delta i}$ the Bessel function of the same order.\\
For a Stark force $F_0$ not satisfying the RET conditions, the coupling to the upper band is strongly suppressed, and almost negligible at least for small particle-particle interband interactions $U_x$. On the other hand if $U_x$ dominates, strong interband coupling is possible even for small forces $F_0$. The latter strongly correlated regime of two energy bands is extremely hard to deal with, especially if one is searching for analytical predictions for the interband dynamics. The results shown in Fig.~\ref{fig:rabi} are just a small step in this direction. In the limit of small atom-atom interactions, the observed collapse and revival times can be determined analytically in good approximation. We quantify small interactions by artificially rescaling the parameters $U_a, U_b, U_x$, which would be obtained by a given scattering length and a given depth of the optical lattice potential \cite{Bloch2008}, by a constant factor $0 < \alpha <1$. For the results of Fig.~\ref{fig:rabi}, $\alpha$ was chosen to be zero (black dotted line), 0.2 (faint red line) and 0.5 (green thick line). The analogy with the Rabi oscillation problem even carries over to those values of interaction strength, since we observe a collapse and later on a revival of the periodic oscillation of the population. Collapse and revival time scale inversely proportional with the strength factor $\alpha$, as shown in Fig.~\ref{fig:scaling}, where the revival time is well approximated by the formula derived in~\cite{Ploetz2010} 
\begin{equation}
\frac{ t_{\rm revival} } { T_{\rm Bloch} } \approx \frac{2F_0}{\alpha U_x J_0^2\left(\frac{J_a}{F_0} \right)J_0^2\left(\frac{J_b}{F_0} \right)}\,,
\label{eq:rev}
\end{equation}
with the zeroth order Bessel function $J_0$.
This formula arises from a perturbative calculation of the effect of atom-atom interactions for small $\alpha U_{a,b,x}\ll F_0$ starting from the single-particle solution, which itself is known within the effective two-state model, and assuming a delocalized initial state along the lattice. From Eq.~(\ref{eq:rev}) the collapse time was estimated in~ \cite{Meystre2007} as $t_{\rm collapse}\approx t_{\rm revival}/(\pi \sigma_s )$, with the effective number $\sigma_s$ of additionally coupled many-particle states as compared to the single-particle two-state model. This collapse is analogous to that of the  Rabi oscillations in the presence of atomic interactions, or to the collapse arising whenever the phase evolution of each $s$ basis state is nonlinear in the particle number. Notice that the collapse and revival phenomena of Fig.~\ref{fig:rabi} stem from a degradation (arising from interactions) of single particle {\em interband} oscillations (with original period given by Eq.~(\ref{eq:osc}) which just depends on the force $F_0$). So, even if there are analogies to the  collapses and revivals observed in BEC~\cite{Greiner2002,Anderlini2006,SebbyStrabley2007,Will2010}, their  origins are different. In the BEC  investigations the collapse-revival oscillations were produced by the interaction within a single-band (in~\cite{Will2010} by atomic interactions depending on higher power of the well occupation number). Therefore those oscillations would {\em not} at all occur when the lower band nonlinear interaction(s) is (are) suppressed, equivalent to $U_a=0$ in the model here discussed.\\
\indent The above steps may be expanded in different directions within the realm of true many-body dynamics and tunneling, with great perspectives for many-body induced RET effects. Remaining questions are, for instance, the study of the strongly correlated regime of strong particle and strong interband interactions simultaneously, and the enlargement of our closed two band model in order to allow for a realistic description of experiments similar to the ones reported in Section~\ref{sec4_2} now carried over into the realm of strong many-body interactions.

\section{\label{sec6} Conclusions and perspectives on RET}
   This chapter has presented and discussed the RET investigations performed with cold and ultracold atoms. Owing to the reached high level of control on the atom initial preparation and on the realization of potentials with arbitrary shapes, the atomic physics community has reproduced and analyzed basic quantum mechanics phenomena well established, and with important applications, within the solid state physics community. An important feature associated to the investigations on the atoms, compared to those on electrons in a solid,  is the absence of decoherence phenomena. Therefore quantum interference phenomena may play an enormous role on  the tunneling temporal evolution of the cold atoms. For the ultracold atoms an additional characteristic  is the presence of interatomic interactions, that modify the position of the energy levels and therefore greatly influence the RET.  In more complex configurations the atomic interactions lead to a very complex Hamiltonian whose action on the atoms requires large computational efforts or analyses based on perturbation approaches.\\
   \indent Our analysis was restricted to potentials which are either not explicitly time-dependent or lead to a temporal evolution of the atomic wavefunction  corresponding to an adiabatic evolution of the atomic system. Tunneling processes produced by a non-adiabatic atomic evolution are described in other chapters of this volume. \\
   \indent  Macroscopic quantum tunneling is an important direction of research well investigated by the solid state physics community. Up to now no clear evidence of that tunneling was reported by the BEC community even if configurations for the occurrence of macroscopic quantum tunneling in Bose-Einstein condensates have  been proposed by different authors. Ueda and Leggett~\cite{Ueda1998,UedaReply1998} examined the instability of a collective mode in a BEC with attractive interaction induced by macroscopic tunneling. Thus,  a collective variable the spatial width of BEC is analyzed a as a tunneling variable.  Carr {\em et al.}~\cite{Carr2005}  studied BEC in a potential of finite depth, harmonic for small radii and decaying as a Gaussian for large radii, which supports both bound and quasi-bound states. The atomic nonlinearity  transforming bound states into quasi-bound ones, leads to macroscopic quantum tunneling.  The experimental observation of such macroscopic tunneling would enlarge the quantum simulation configurations explored with ultracold atoms. 

\begin{acknowledgments}
E.A. thanks the IFRAF, Paris (France), for the financial support at the \'Ecole Normale Sup\'erieure where this work was initiated. 
We gratefully acknowledge also support from the Project NAMEQUAM of the Future and Emerging Technologies (FET) programme 
within the Seventh Framework Programme for Research of the European Commission (FET-Open grant number: 225187), 
the PRIN Project of the MIUR of Italy, and the Excellence Initiative by the German Research Foundation (DFG) through the 
Heidelberg Graduate School of Fundamental Physics (Grant No. GSC 129/1), the Frontier Innovation Fonds and 
the Global Networks Mobility Measures of the University of Heidelberg. S.W. is especially grateful to the Heidelberg Academy of Sciences and Humanities for the Academy Award 2010 and to the Hengstberger Foundation for support by the Klaus-Georg and Sigrid Hengstberger Prize 2009. 
This review was stimulated by the tunneling experiments performed in Pisa by D. Ciampini,  H. Lignier, O. Morsch C. Sias, and  
A. Zenesini, and we thank all of them for the continuous valuable discussions. Finally, we would like to thank our theory collaborators, 
G. Tayebirad, N. L\"orch, A. Tomadin, P. Schlagheck, A. Kolovsky, P. Pl\"otz, D. Witthaut, J. Madro\~nero, and R. Mannella for their help in pushing forward this work.
\end{acknowledgments}
 



\begin{thebibliography}{98}
\expandafter\ifx\csname natexlab\endcsname\relax\def\natexlab#1{#1}\fi
\expandafter\ifx\csname bibnamefont\endcsname\relax
  \def\bibnamefont#1{#1}\fi
\expandafter\ifx\csname bibfnamefont\endcsname\relax
  \def\bibfnamefont#1{#1}\fi
\expandafter\ifx\csname citenamefont\endcsname\relax
  \def\citenamefont#1{#1}\fi
\expandafter\ifx\csname url\endcsname\relax
  \def\url#1{\texttt{#1}}\fi
\expandafter\ifx\csname urlprefix\endcsname\relax\def\urlprefix{URL }\fi
\providecommand{\bibinfo}[2]{#2}
\providecommand{\eprint}[2][]{\url{#2}}

\bibitem[{\citenamefont{Davis and Heller}(1981)}]{DavisHeller1981}
\bibinfo{author}{\bibfnamefont{M.~J.} \bibnamefont{Davis}} \bibnamefont{and}
  \bibinfo{author}{\bibfnamefont{E.~J.} \bibnamefont{Heller}},
  \bibinfo{journal}{J. Chem. Phys.} \textbf{\bibinfo{volume}{75}},
  \bibinfo{pages}{246} (\bibinfo{year}{1981}).

\bibitem[{\citenamefont{Tsu and Esaki}(1973)}]{Tsu1973}
\bibinfo{author}{\bibfnamefont{R.}~\bibnamefont{Tsu}} \bibnamefont{and}
  \bibinfo{author}{\bibfnamefont{L.}~\bibnamefont{Esaki}},
  \bibinfo{journal}{Appl.Phys. Lett.} \textbf{\bibinfo{volume}{22}},
  \bibinfo{pages}{562} (\bibinfo{year}{1973}).

\bibitem[{\citenamefont{Chang et~al.}(1974)\citenamefont{Chang, Esaki, and
  Tsu}}]{Chang1974}
\bibinfo{author}{\bibfnamefont{L.~L.} \bibnamefont{Chang}},
  \bibinfo{author}{\bibfnamefont{L.}~\bibnamefont{Esaki}}, \bibnamefont{and}
  \bibinfo{author}{\bibfnamefont{R.}~\bibnamefont{Tsu}},
  \bibinfo{journal}{Appl. Phys. Lett.} \textbf{\bibinfo{volume}{24}},
  \bibinfo{pages}{593} (\bibinfo{year}{1974}).

\bibitem[{\citenamefont{Chang et~al.}(1991)\citenamefont{Chang, Mendez, and
  [Eds.]}}]{Chang1991}
\bibinfo{author}{\bibfnamefont{L.}~\bibnamefont{Chang}},
  \bibinfo{author}{\bibfnamefont{E.}~\bibnamefont{Mendez}}, \bibnamefont{and}
  \bibinfo{author}{\bibfnamefont{C.~T.} \bibnamefont{[Eds.]}},
  \emph{\bibinfo{title}{Resonant Tunneling in Semiconductors}}
  (\bibinfo{publisher}{Plenum Press}, \bibinfo{address}{Amstedam},
  \bibinfo{year}{1991}).

\bibitem[{\citenamefont{Esaki}(1986)}]{Esaki1986}
\bibinfo{author}{\bibfnamefont{L.}~\bibnamefont{Esaki}}, \bibinfo{journal}{IEEE
  Journal Quant. Electr.} \textbf{\bibinfo{volume}{QE-22}},
  \bibinfo{pages}{1611} (\bibinfo{year}{1986}).

\bibitem[{\citenamefont{Leo}(2003)}]{Leo2003}
\bibinfo{author}{\bibfnamefont{K.}~\bibnamefont{Leo}},
  \emph{\bibinfo{title}{High-Field Transport in Semiconductor Superlattices}}
  (\bibinfo{publisher}{Springer Verlag}, \bibinfo{address}{Berlin},
  \bibinfo{year}{2003}).

\bibitem[{\citenamefont{Glutsch}(2004)}]{Glutsch2004}
\bibinfo{author}{\bibfnamefont{S.}~\bibnamefont{Glutsch}},
  \bibinfo{journal}{Phys. Rev. B} \textbf{\bibinfo{volume}{69}},
  \bibinfo{pages}{235317} (\bibinfo{year}{2004}).

\bibitem[{\citenamefont{Wagner and Mizuta}(1993)}]{Wagner1993}
\bibinfo{author}{\bibfnamefont{M.}~\bibnamefont{Wagner}} \bibnamefont{and}
  \bibinfo{author}{\bibfnamefont{H.}~\bibnamefont{Mizuta}},
  \bibinfo{journal}{Phys. Rev. B} \textbf{\bibinfo{volume}{48}},
  \bibinfo{pages}{14393} (\bibinfo{year}{1993}).

\bibitem[{\citenamefont{Rosam et~al.}(2003)\citenamefont{Rosam, Leo, Gl\"uck,
  Keck, Korsch, Zimmer, and K\"ohler}}]{Rosam2003}
\bibinfo{author}{\bibfnamefont{B.}~\bibnamefont{Rosam}},
  \bibinfo{author}{\bibfnamefont{K.}~\bibnamefont{Leo}},
  \bibinfo{author}{\bibfnamefont{M.}~\bibnamefont{Gl\"uck}},
  \bibinfo{author}{\bibfnamefont{F.}~\bibnamefont{Keck}},
  \bibinfo{author}{\bibfnamefont{H.~J.} \bibnamefont{Korsch}},
  \bibinfo{author}{\bibfnamefont{F.}~\bibnamefont{Zimmer}}, \bibnamefont{and}
  \bibinfo{author}{\bibfnamefont{K.}~\bibnamefont{K\"ohler}},
  \bibinfo{journal}{Phys. Rev. B} \textbf{\bibinfo{volume}{68}},
  \bibinfo{pages}{125301} (\bibinfo{year}{2003}).

\bibitem[{\citenamefont{Mizuta and Tanoue}(1995)}]{Mizuta1995}
\bibinfo{author}{\bibfnamefont{H.}~\bibnamefont{Mizuta}} \bibnamefont{and}
  \bibinfo{author}{\bibfnamefont{T.}~\bibnamefont{Tanoue}},
  \emph{\bibinfo{title}{The Physics and Applications of Resonant Tunnelling
  Diodes}} (\bibinfo{publisher}{Cambridge Univ. Press},
  \bibinfo{address}{Cambridge, UK}, \bibinfo{year}{1995}).

\bibitem[{\citenamefont{Ben~Dahan et~al.}(1996)\citenamefont{Ben~Dahan, Peik,
  Reichel, Castin, and Salomon}}]{Peik1996}
\bibinfo{author}{\bibfnamefont{M.}~\bibnamefont{Ben~Dahan}},
  \bibinfo{author}{\bibfnamefont{E.}~\bibnamefont{Peik}},
  \bibinfo{author}{\bibfnamefont{J.}~\bibnamefont{Reichel}},
  \bibinfo{author}{\bibfnamefont{Y.}~\bibnamefont{Castin}}, \bibnamefont{and}
  \bibinfo{author}{\bibfnamefont{C.}~\bibnamefont{Salomon}},
  \bibinfo{journal}{Phys. Rev. Lett.} \textbf{\bibinfo{volume}{76}},
  \bibinfo{pages}{4508} (\bibinfo{year}{1996}).

\bibitem[{\citenamefont{Raizen et~al.}(1997)\citenamefont{Raizen, Salomon, and
  Niu}}]{Raizen1997}
\bibinfo{author}{\bibfnamefont{M.}~\bibnamefont{Raizen}},
  \bibinfo{author}{\bibfnamefont{C.}~\bibnamefont{Salomon}}, \bibnamefont{and}
  \bibinfo{author}{\bibfnamefont{Q.}~\bibnamefont{Niu}},
  \bibinfo{journal}{Physics Today} \textbf{\bibinfo{volume}{50}},
  \bibinfo{pages}{30} (\bibinfo{year}{1997}).

\bibitem[{\citenamefont{Roati et~al.}(2004)\citenamefont{Roati, de~Mirandes,
  Ferlaino, Ott, Modugno, and Inguscio}}]{Roati2004}
\bibinfo{author}{\bibfnamefont{G.}~\bibnamefont{Roati}},
  \bibinfo{author}{\bibfnamefont{E.}~\bibnamefont{de~Mirandes}},
  \bibinfo{author}{\bibfnamefont{F.}~\bibnamefont{Ferlaino}},
  \bibinfo{author}{\bibfnamefont{H.}~\bibnamefont{Ott}},
  \bibinfo{author}{\bibfnamefont{G.}~\bibnamefont{Modugno}}, \bibnamefont{and}
  \bibinfo{author}{\bibfnamefont{M.}~\bibnamefont{Inguscio}},
  \bibinfo{journal}{Phys. Rev. Lett.} \textbf{\bibinfo{volume}{92}},
  \bibinfo{pages}{230402} (\bibinfo{year}{2004}).

\bibitem[{\citenamefont{Morsch and Oberthaler}(2006)}]{Morsch2006}
\bibinfo{author}{\bibfnamefont{O.}~\bibnamefont{Morsch}} \bibnamefont{and}
  \bibinfo{author}{\bibfnamefont{M.}~\bibnamefont{Oberthaler}},
  \bibinfo{journal}{Rev. Mod. Phys.} \textbf{\bibinfo{volume}{78}},
  \bibinfo{eid}{179} (\bibinfo{year}{2006}).

\bibitem[{\citenamefont{Gustavsson et~al.}(2008)\citenamefont{Gustavsson,
  Haller, Mark, Danzl, Rojas-Kopeinig, and N\"agerl}}]{Gustavsson2008}
\bibinfo{author}{\bibfnamefont{M.}~\bibnamefont{Gustavsson}},
  \bibinfo{author}{\bibfnamefont{E.}~\bibnamefont{Haller}},
  \bibinfo{author}{\bibfnamefont{M.~J.} \bibnamefont{Mark}},
  \bibinfo{author}{\bibfnamefont{J.~G.} \bibnamefont{Danzl}},
  \bibinfo{author}{\bibfnamefont{G.}~\bibnamefont{Rojas-Kopeinig}},
  \bibnamefont{and} \bibinfo{author}{\bibfnamefont{H.-C.}
  \bibnamefont{N\"agerl}}, \bibinfo{journal}{Phys. Rev. Lett.}
  \textbf{\bibinfo{volume}{100}}, \bibinfo{pages}{080404}
  (\bibinfo{year}{2008}).

\bibitem[{\citenamefont{Bloch et~al.}(2008)\citenamefont{Bloch, Dalibard, and
  Zwerger}}]{Bloch2008}
\bibinfo{author}{\bibfnamefont{I.}~\bibnamefont{Bloch}},
  \bibinfo{author}{\bibfnamefont{J.}~\bibnamefont{Dalibard}}, \bibnamefont{and}
  \bibinfo{author}{\bibfnamefont{W.}~\bibnamefont{Zwerger}},
  \bibinfo{journal}{Rev. Mod. Phys.} \textbf{\bibinfo{volume}{80}},
  \bibinfo{eid}{885} (\bibinfo{year}{2008}).

\bibitem[{\citenamefont{Teo et~al.}(2002)\citenamefont{Teo, Guest, and
  Raithel}}]{Teo2002}
\bibinfo{author}{\bibfnamefont{B.~K.} \bibnamefont{Teo}},
  \bibinfo{author}{\bibfnamefont{J.~R.} \bibnamefont{Guest}}, \bibnamefont{and}
  \bibinfo{author}{\bibfnamefont{G.}~\bibnamefont{Raithel}},
  \bibinfo{journal}{Phys. Rev. Lett.} \textbf{\bibinfo{volume}{88}},
  \bibinfo{pages}{173001} (\bibinfo{year}{2002}).

\bibitem[{\citenamefont{Greiner et~al.}(2002)\citenamefont{Greiner, Mandel,
  Esslinger, H{\"a}nsch, and Bloch}}]{Greiner2002}
\bibinfo{author}{\bibfnamefont{M.}~\bibnamefont{Greiner}},
  \bibinfo{author}{\bibfnamefont{O.}~\bibnamefont{Mandel}},
  \bibinfo{author}{\bibfnamefont{T.}~\bibnamefont{Esslinger}},
  \bibinfo{author}{\bibfnamefont{T.~W.} \bibnamefont{H{\"a}nsch}},
  \bibnamefont{and} \bibinfo{author}{\bibfnamefont{I.}~\bibnamefont{Bloch}},
  \bibinfo{journal}{Nature} \textbf{\bibinfo{volume}{415}}, \bibinfo{pages}{39}
  (\bibinfo{year}{2002}).

\bibitem[{\citenamefont{F\"olling et~al.}(2007)\citenamefont{F\"olling,
  Trotzky, Cheinet, Feld, Saers, Widera, M\"uller, and Bloch}}]{Foelling2007}
\bibinfo{author}{\bibfnamefont{S.}~\bibnamefont{F\"olling}},
  \bibinfo{author}{\bibfnamefont{S.}~\bibnamefont{Trotzky}},
  \bibinfo{author}{\bibfnamefont{P.}~\bibnamefont{Cheinet}},
  \bibinfo{author}{\bibfnamefont{M.}~\bibnamefont{Feld}},
  \bibinfo{author}{\bibfnamefont{R.}~\bibnamefont{Saers}},
  \bibinfo{author}{\bibfnamefont{A.}~\bibnamefont{Widera}},
  \bibinfo{author}{\bibfnamefont{T.}~\bibnamefont{M\"uller}}, \bibnamefont{and}
  \bibinfo{author}{\bibfnamefont{I.}~\bibnamefont{Bloch}},
  \bibinfo{journal}{Nature} \textbf{\bibinfo{volume}{448}},
  \bibinfo{pages}{1029} (\bibinfo{year}{2007}).

\bibitem[{\citenamefont{Wilkinson et~al.}(1996)\citenamefont{Wilkinson,
  Bharucha, Madison, Niu, and Raizen}}]{Wilkinson1996}
\bibinfo{author}{\bibfnamefont{S.~R.} \bibnamefont{Wilkinson}},
  \bibinfo{author}{\bibfnamefont{C.~F.} \bibnamefont{Bharucha}},
  \bibinfo{author}{\bibfnamefont{K.~W.} \bibnamefont{Madison}},
  \bibinfo{author}{\bibfnamefont{Q.}~\bibnamefont{Niu}}, \bibnamefont{and}
  \bibinfo{author}{\bibfnamefont{M.~G.} \bibnamefont{Raizen}},
  \bibinfo{journal}{Phys. Rev. Lett.} \textbf{\bibinfo{volume}{76}},
  \bibinfo{pages}{4512} (\bibinfo{year}{1996}).

\bibitem[{\citenamefont{K\"ohl et~al.}(2005)\citenamefont{K\"ohl, Moritz,
  St\"oferle, G\"unter, and Esslinger}}]{Koehl2005}
\bibinfo{author}{\bibfnamefont{M.}~\bibnamefont{K\"ohl}},
  \bibinfo{author}{\bibfnamefont{H.}~\bibnamefont{Moritz}},
  \bibinfo{author}{\bibfnamefont{T.}~\bibnamefont{St\"oferle}},
  \bibinfo{author}{\bibfnamefont{K.}~\bibnamefont{G\"unter}}, \bibnamefont{and}
  \bibinfo{author}{\bibfnamefont{T.}~\bibnamefont{Esslinger}},
  \bibinfo{journal}{Phys. Rev. Lett.} \textbf{\bibinfo{volume}{94}},
  \bibinfo{pages}{080403} (\bibinfo{year}{2005}).

\bibitem[{\citenamefont{Lee et~al.}(2007)\citenamefont{Lee, Ostrovskaya, and
  Kivshar}}]{Lee2007}
\bibinfo{author}{\bibfnamefont{C.}~\bibnamefont{Lee}},
  \bibinfo{author}{\bibfnamefont{E.~A.} \bibnamefont{Ostrovskaya}},
  \bibnamefont{and} \bibinfo{author}{\bibfnamefont{Y.}~\bibnamefont{Kivshar}},
  \bibinfo{journal}{J. Phys. B: At. Molec. Opt. Phys.}
  \textbf{\bibinfo{volume}{40}}, \bibinfo{pages}{4235} (\bibinfo{year}{2007}).

\bibitem[{\citenamefont{Sias et~al.}(2007)\citenamefont{Sias, Zenesini,
  Lignier, Wimberger, Ciampini, Morsch, and Arimondo}}]{Sias2007}
\bibinfo{author}{\bibfnamefont{C.}~\bibnamefont{Sias}},
  \bibinfo{author}{\bibfnamefont{A.}~\bibnamefont{Zenesini}},
  \bibinfo{author}{\bibfnamefont{H.}~\bibnamefont{Lignier}},
  \bibinfo{author}{\bibfnamefont{S.}~\bibnamefont{Wimberger}},
  \bibinfo{author}{\bibfnamefont{D.}~\bibnamefont{Ciampini}},
  \bibinfo{author}{\bibfnamefont{O.}~\bibnamefont{Morsch}}, \bibnamefont{and}
  \bibinfo{author}{\bibfnamefont{E.}~\bibnamefont{Arimondo}},
  \bibinfo{journal}{Phys. Rev. Lett.} \textbf{\bibinfo{volume}{98}},
  \bibinfo{eid}{120403} (\bibinfo{year}{2007}).

\bibitem[{\citenamefont{Zenesini et~al.}(2008)\citenamefont{Zenesini, Sias,
  Lignier, Singh, Ciampini, Morsch, Mannella, Arimondo, Tomadin, and
  Wimberger}}]{Zenesini2008}
\bibinfo{author}{\bibfnamefont{A.}~\bibnamefont{Zenesini}},
  \bibinfo{author}{\bibfnamefont{C.}~\bibnamefont{Sias}},
  \bibinfo{author}{\bibfnamefont{H.}~\bibnamefont{Lignier}},
  \bibinfo{author}{\bibfnamefont{Y.}~\bibnamefont{Singh}},
  \bibinfo{author}{\bibfnamefont{D.}~\bibnamefont{Ciampini}},
  \bibinfo{author}{\bibfnamefont{O.}~\bibnamefont{Morsch}},
  \bibinfo{author}{\bibfnamefont{R.}~\bibnamefont{Mannella}},
  \bibinfo{author}{\bibfnamefont{E.}~\bibnamefont{Arimondo}},
  \bibinfo{author}{\bibfnamefont{A.}~\bibnamefont{Tomadin}}, \bibnamefont{and}
  \bibinfo{author}{\bibfnamefont{S.}~\bibnamefont{Wimberger}},
  \bibinfo{journal}{NJP} \textbf{\bibinfo{volume}{10}},
  \bibinfo{pages}{0530388} (\bibinfo{year}{2008}).

\bibitem[{\citenamefont{Cristiani et~al.}(2002)\citenamefont{Cristiani, Morsch,
  M\"uller, Ciampini, and Arimondo}}]{Cristiani2002}
\bibinfo{author}{\bibfnamefont{M.}~\bibnamefont{Cristiani}},
  \bibinfo{author}{\bibfnamefont{O.}~\bibnamefont{Morsch}},
  \bibinfo{author}{\bibfnamefont{J.~H.} \bibnamefont{M\"uller}},
  \bibinfo{author}{\bibfnamefont{D.}~\bibnamefont{Ciampini}}, \bibnamefont{and}
  \bibinfo{author}{\bibfnamefont{E.}~\bibnamefont{Arimondo}},
  \bibinfo{journal}{Phys. Rev. A} \textbf{\bibinfo{volume}{65}},
  \bibinfo{pages}{063612} (\bibinfo{year}{2002}).

\bibitem[{\citenamefont{Grimm et~al.}(2000)\citenamefont{Grimm,
  Weidem{\"u}ller, and Ovchinnikov}}]{Grimm2000}
\bibinfo{author}{\bibfnamefont{R.}~\bibnamefont{Grimm}},
  \bibinfo{author}{\bibfnamefont{M.}~\bibnamefont{Weidem{\"u}ller}},
  \bibnamefont{and} \bibinfo{author}{\bibfnamefont{Y.~B.}
  \bibnamefont{Ovchinnikov}}, \bibinfo{journal}{Adv. At. Mol. Opt. Phys.}
  \textbf{\bibinfo{volume}{42}}, \bibinfo{pages}{95} (\bibinfo{year}{2000}).

\bibitem[{\citenamefont{Nenciu}(1991)}]{Nenciu1991}
\bibinfo{author}{\bibfnamefont{G.}~\bibnamefont{Nenciu}},
  \bibinfo{journal}{Rev. Mod. Phys.} \textbf{\bibinfo{volume}{63}},
  \bibinfo{pages}{91} (\bibinfo{year}{1991}).

\bibitem[{\citenamefont{Gl\"uck et~al.}(2002)\citenamefont{Gl\"uck, Kolovsky,
  and Korsch}}]{Glueck2002}
\bibinfo{author}{\bibfnamefont{M.}~\bibnamefont{Gl\"uck}},
  \bibinfo{author}{\bibfnamefont{A.~R.} \bibnamefont{Kolovsky}},
  \bibnamefont{and} \bibinfo{author}{\bibfnamefont{H.~J.}
  \bibnamefont{Korsch}}, \bibinfo{journal}{Physics Reports}
  \textbf{\bibinfo{volume}{366}}, \bibinfo{pages}{103 } (\bibinfo{year}{2002}).

\bibitem[{\citenamefont{Holthaus}(2000)}]{Holthaus2000}
\bibinfo{author}{\bibfnamefont{M.}~\bibnamefont{Holthaus}},
  \bibinfo{journal}{J. Opt. B: Quantum Semicl. Opt.}
  \textbf{\bibinfo{volume}{2}}, \bibinfo{pages}{589} (\bibinfo{year}{2000}).

\bibitem[{\citenamefont{Landau}(1932)}]{Landau1932}
\bibinfo{author}{\bibfnamefont{L.}~\bibnamefont{Landau}},
  \bibinfo{journal}{Phys. Z. Sowjetunion} \textbf{\bibinfo{volume}{2}},
  \bibinfo{pages}{46} (\bibinfo{year}{1932}).

\bibitem[{\citenamefont{Zener}(1932)}]{Zener1932}
\bibinfo{author}{\bibfnamefont{C.}~\bibnamefont{Zener}},
  \bibinfo{journal}{Proc. R. Soc. London, Ser. A}
  \textbf{\bibinfo{volume}{137}}, \bibinfo{pages}{696} (\bibinfo{year}{1932}).

\bibitem[{\citenamefont{Aschcroft and Mermin}(1976)}]{Ashcroft1976}
\bibinfo{author}{\bibfnamefont{N.}~\bibnamefont{Aschcroft}} \bibnamefont{and}
  \bibinfo{author}{\bibfnamefont{N.}~\bibnamefont{Mermin}},
  \emph{\bibinfo{title}{Solid state physics}} (\bibinfo{publisher}{Saunders
  College}, \bibinfo{address}{Philadelphia}, \bibinfo{year}{1976}).

\bibitem[{\citenamefont{Zwerger}(2003)}]{Zwerger2003}
\bibinfo{author}{\bibfnamefont{W.}~\bibnamefont{Zwerger}}, \bibinfo{journal}{J.
  Opt. B: Quantum Semicl. Opt.} \textbf{\bibinfo{volume}{5}},
  \bibinfo{pages}{S9} (\bibinfo{year}{2003}).

\bibitem[{\citenamefont{Pethick and Smith}(2002)}]{Pethick2002}
\bibinfo{author}{\bibfnamefont{C.~J.} \bibnamefont{Pethick}} \bibnamefont{and}
  \bibinfo{author}{\bibfnamefont{H.}~\bibnamefont{Smith}},
  \emph{\bibinfo{title}{Bose-{E}instein condensation in dilute gases}}
  (\bibinfo{publisher}{Cambridge Univ. Press}, \bibinfo{address}{Cambridge,
  UK}, \bibinfo{year}{2002}).

\bibitem[{\citenamefont{Pitaevskii and Stringari}(2003)}]{Pitaevskii2003}
\bibinfo{author}{\bibfnamefont{L.}~\bibnamefont{Pitaevskii}} \bibnamefont{and}
  \bibinfo{author}{\bibfnamefont{S.}~\bibnamefont{Stringari}},
  \emph{\bibinfo{title}{Bose-{E}instein condensation}}
  (\bibinfo{publisher}{Oxford Univ. Press}, \bibinfo{address}{Oxford, UK},
  \bibinfo{year}{2003}).

\bibitem[{\citenamefont{Castin et~al.}(1994)\citenamefont{Castin,
  Berg-S\"orensen, Dalibard, and M\"olmer}}]{Castin1994}
\bibinfo{author}{\bibfnamefont{Y.}~\bibnamefont{Castin}},
  \bibinfo{author}{\bibfnamefont{K.}~\bibnamefont{Berg-S\"orensen}},
  \bibinfo{author}{\bibfnamefont{J.}~\bibnamefont{Dalibard}}, \bibnamefont{and}
  \bibinfo{author}{\bibfnamefont{K.}~\bibnamefont{M\"olmer}},
  \bibinfo{journal}{Phys. Rev. A} \textbf{\bibinfo{volume}{50}},
  \bibinfo{pages}{5092} (\bibinfo{year}{1994}).

\bibitem[{\citenamefont{Dutta et~al.}(1999)\citenamefont{Dutta, Teo, and
  Raithel}}]{Dutta1999}
\bibinfo{author}{\bibfnamefont{S.~K.} \bibnamefont{Dutta}},
  \bibinfo{author}{\bibfnamefont{B.~K.} \bibnamefont{Teo}}, \bibnamefont{and}
  \bibinfo{author}{\bibfnamefont{G.}~\bibnamefont{Raithel}},
  \bibinfo{journal}{Phys. Rev. Lett.} \textbf{\bibinfo{volume}{83}},
  \bibinfo{pages}{1934} (\bibinfo{year}{1999}).

\bibitem[{\citenamefont{Haycock et~al.}(2000)\citenamefont{Haycock, Alsing,
  Deutsch, Grondalski, and Jessen}}]{Haycock2000}
\bibinfo{author}{\bibfnamefont{D.~L.} \bibnamefont{Haycock}},
  \bibinfo{author}{\bibfnamefont{P.~M.} \bibnamefont{Alsing}},
  \bibinfo{author}{\bibfnamefont{I.~H.} \bibnamefont{Deutsch}},
  \bibinfo{author}{\bibfnamefont{J.}~\bibnamefont{Grondalski}},
  \bibnamefont{and} \bibinfo{author}{\bibfnamefont{P.~S.}
  \bibnamefont{Jessen}}, \bibinfo{journal}{Phys. Rev. Lett.}
  \textbf{\bibinfo{volume}{85}}, \bibinfo{pages}{3365} (\bibinfo{year}{2000}).

\bibitem[{\citenamefont{Kierig et~al.}(2008)\citenamefont{Kierig,
  Schnorrberger, Schietinger, Tomkovic, and Oberthaler}}]{Kierig2008}
\bibinfo{author}{\bibfnamefont{E.}~\bibnamefont{Kierig}},
  \bibinfo{author}{\bibfnamefont{U.}~\bibnamefont{Schnorrberger}},
  \bibinfo{author}{\bibfnamefont{A.}~\bibnamefont{Schietinger}},
  \bibinfo{author}{\bibfnamefont{J.}~\bibnamefont{Tomkovic}}, \bibnamefont{and}
  \bibinfo{author}{\bibfnamefont{M.~K.} \bibnamefont{Oberthaler}},
  \bibinfo{journal}{Phys. Rev. Lett.} \textbf{\bibinfo{volume}{100}},
  \bibinfo{pages}{190405} (\bibinfo{year}{2008}).

\bibitem[{\citenamefont{Dounas-Frazer et~al.}(2007)\citenamefont{Dounas-Frazer,
  Hermundstad, and Carr}}]{DounasFrazer2007}
\bibinfo{author}{\bibfnamefont{D.~R.} \bibnamefont{Dounas-Frazer}},
  \bibinfo{author}{\bibfnamefont{A.~M.} \bibnamefont{Hermundstad}},
  \bibnamefont{and} \bibinfo{author}{\bibfnamefont{L.~D.} \bibnamefont{Carr}},
  \bibinfo{journal}{Phys. Rev. Lett.} \textbf{\bibinfo{volume}{99}},
  \bibinfo{eid}{200402} (\bibinfo{year}{2007}).

\bibitem[{\citenamefont{Albiez et~al.}(2005)\citenamefont{Albiez, Gati,
  F\"olling, Hunsmann, Cristiani, and Oberthaler}}]{Albiez2005}
\bibinfo{author}{\bibfnamefont{M.}~\bibnamefont{Albiez}},
  \bibinfo{author}{\bibfnamefont{R.}~\bibnamefont{Gati}},
  \bibinfo{author}{\bibfnamefont{J.}~\bibnamefont{F\"olling}},
  \bibinfo{author}{\bibfnamefont{S.}~\bibnamefont{Hunsmann}},
  \bibinfo{author}{\bibfnamefont{M.}~\bibnamefont{Cristiani}},
  \bibnamefont{and} \bibinfo{author}{\bibfnamefont{M.~K.}
  \bibnamefont{Oberthaler}}, \bibinfo{journal}{Phys. Rev. Lett.}
  \textbf{\bibinfo{volume}{95}}, \bibinfo{pages}{010402}
  (\bibinfo{year}{2005}).

\bibitem[{\citenamefont{Khomeriki et~al.}(2006)\citenamefont{Khomeriki, Ruffo,
  and Wimberger}}]{Khomeriki2006}
\bibinfo{author}{\bibfnamefont{R.}~\bibnamefont{Khomeriki}},
  \bibinfo{author}{\bibfnamefont{S.}~\bibnamefont{Ruffo}}, \bibnamefont{and}
  \bibinfo{author}{\bibfnamefont{S.}~\bibnamefont{Wimberger}},
  \bibinfo{journal}{Europhys. Lett.} \textbf{\bibinfo{volume}{77}},
  \bibinfo{pages}{40005} (\bibinfo{year}{2006}).

\bibitem[{\citenamefont{Averbukh et~al.}(2002)\citenamefont{Averbukh, Osovski,
  and Moiseyev}}]{Averbukh2002}
\bibinfo{author}{\bibfnamefont{V.}~\bibnamefont{Averbukh}},
  \bibinfo{author}{\bibfnamefont{S.}~\bibnamefont{Osovski}}, \bibnamefont{and}
  \bibinfo{author}{\bibfnamefont{N.}~\bibnamefont{Moiseyev}},
  \bibinfo{journal}{Phys. Rev. Lett.} \textbf{\bibinfo{volume}{89}},
  \bibinfo{pages}{253201} (\bibinfo{year}{2002}).

\bibitem[{\citenamefont{Hensinger et~al.}(2004)\citenamefont{Hensinger,
  Mouchet, Julienne, Delande, Heckenberg, and
  Rubinsztein-Dunlop}}]{Hensinger2004}
\bibinfo{author}{\bibfnamefont{W.~K.} \bibnamefont{Hensinger}},
  \bibinfo{author}{\bibfnamefont{A.}~\bibnamefont{Mouchet}},
  \bibinfo{author}{\bibfnamefont{P.~S.} \bibnamefont{Julienne}},
  \bibinfo{author}{\bibfnamefont{D.}~\bibnamefont{Delande}},
  \bibinfo{author}{\bibfnamefont{N.~R.} \bibnamefont{Heckenberg}},
  \bibnamefont{and}
  \bibinfo{author}{\bibfnamefont{H.}~\bibnamefont{Rubinsztein-Dunlop}},
  \bibinfo{journal}{Phys. Rev. A} \textbf{\bibinfo{volume}{70}},
  \bibinfo{pages}{013408} (\bibinfo{year}{2004}).

\bibitem[{\citenamefont{Steck et~al.}(2001)\citenamefont{Steck, Oskay, and
  Raizen}}]{Steck2001}
\bibinfo{author}{\bibfnamefont{D.~A.} \bibnamefont{Steck}},
  \bibinfo{author}{\bibfnamefont{W.~H.} \bibnamefont{Oskay}}, \bibnamefont{and}
  \bibinfo{author}{\bibfnamefont{M.~G.} \bibnamefont{Raizen}},
  \bibinfo{journal}{Science} \textbf{\bibinfo{volume}{293}},
  \bibinfo{pages}{274} (\bibinfo{year}{2001}).

\bibitem[{\citenamefont{Hensinger et~al.}(2001)\citenamefont{Hensinger,
  Haffner, Browaeys, Heckenberg, Helmerson, McKenzie, Milburn, Phillips,
  Rolston, Rubinsztein-Dunlop et~al.}}]{Hensinger2001}
\bibinfo{author}{\bibfnamefont{W.}~\bibnamefont{Hensinger}},
  \bibinfo{author}{\bibfnamefont{H.}~\bibnamefont{Haffner}},
  \bibinfo{author}{\bibfnamefont{A.}~\bibnamefont{Browaeys}},
  \bibinfo{author}{\bibfnamefont{N.}~\bibnamefont{Heckenberg}},
  \bibinfo{author}{\bibfnamefont{K.}~\bibnamefont{Helmerson}},
  \bibinfo{author}{\bibfnamefont{C.}~\bibnamefont{McKenzie}},
  \bibinfo{author}{\bibfnamefont{G.}~\bibnamefont{Milburn}},
  \bibinfo{author}{\bibfnamefont{W.}~\bibnamefont{Phillips}},
  \bibinfo{author}{\bibfnamefont{S.}~\bibnamefont{Rolston}},
  \bibinfo{author}{\bibfnamefont{H.}~\bibnamefont{Rubinsztein-Dunlop}},
  \bibnamefont{et~al.}, \bibinfo{journal}{Nature}
  \textbf{\bibinfo{volume}{412}}, \bibinfo{pages}{52} (\bibinfo{year}{2001}).

\bibitem[{\citenamefont{Bohigas et~al.}(1992)\citenamefont{Bohigas, Tomsovic,
  and Ullmo}}]{Bohigas1993}
\bibinfo{author}{\bibfnamefont{O.}~\bibnamefont{Bohigas}},
  \bibinfo{author}{\bibfnamefont{S.}~\bibnamefont{Tomsovic}}, \bibnamefont{and}
  \bibinfo{author}{\bibfnamefont{D.}~\bibnamefont{Ullmo}},
  \bibinfo{journal}{Phys. Rep.} \textbf{\bibinfo{volume}{223}},
  \bibinfo{pages}{43} (\bibinfo{year}{1992}).

\bibitem[{\citenamefont{Dunlap and Kenkre}(1986)}]{Dunlap1986}
\bibinfo{author}{\bibfnamefont{D.~H.} \bibnamefont{Dunlap}} \bibnamefont{and}
  \bibinfo{author}{\bibfnamefont{V.~M.} \bibnamefont{Kenkre}},
  \bibinfo{journal}{Phys.Rev. B} \textbf{\bibinfo{volume}{34}},
  \bibinfo{pages}{3625} (\bibinfo{year}{1986}).

\bibitem[{\citenamefont{Rossi}(1998)}]{Rossi1998}
\bibinfo{author}{\bibfnamefont{F.}~\bibnamefont{Rossi}},
  \bibinfo{journal}{Semicon. Sci. Technol.} \textbf{\bibinfo{volume}{13}},
  \bibinfo{pages}{147} (\bibinfo{year}{1998}).

\bibitem[{\citenamefont{Korsch and Mossmann}(2003)}]{Korsch2003}
\bibinfo{author}{\bibfnamefont{H.~J.} \bibnamefont{Korsch}} \bibnamefont{and}
  \bibinfo{author}{\bibfnamefont{S.}~\bibnamefont{Mossmann}},
  \bibinfo{journal}{Physics Letters A} \textbf{\bibinfo{volume}{317}},
  \bibinfo{pages}{54 } (\bibinfo{year}{2003}).

\bibitem[{\citenamefont{Klumpp et~al.}(2007)\citenamefont{Klumpp, Witthaut, and
  Korsch}}]{Klumpp2007}
\bibinfo{author}{\bibfnamefont{A.}~\bibnamefont{Klumpp}},
  \bibinfo{author}{\bibfnamefont{D.}~\bibnamefont{Witthaut}}, \bibnamefont{and}
  \bibinfo{author}{\bibfnamefont{H.~J.} \bibnamefont{Korsch}},
  \bibinfo{journal}{J. Phys. A: Math. Theor.} \textbf{\bibinfo{volume}{40}},
  \bibinfo{pages}{2299} (\bibinfo{year}{2007}).

\bibitem[{\citenamefont{Tien and Gordon}(1963)}]{Tien1963}
\bibinfo{author}{\bibfnamefont{P.~K.} \bibnamefont{Tien}} \bibnamefont{and}
  \bibinfo{author}{\bibfnamefont{J.~P.} \bibnamefont{Gordon}},
  \bibinfo{journal}{Phys. Rev.} \textbf{\bibinfo{volume}{129}},
  \bibinfo{pages}{647} (\bibinfo{year}{1963}).

\bibitem[{\citenamefont{Keay et~al.}(1995{\natexlab{a}})\citenamefont{Keay,
  Allen, Gal\'an, Kaminski, Campman, Gossard, Bhattacharya, and
  Rodwell}}]{Keay1995}
\bibinfo{author}{\bibfnamefont{B.~J.} \bibnamefont{Keay}},
  \bibinfo{author}{\bibfnamefont{S.~J.} \bibnamefont{Allen}},
  \bibinfo{author}{\bibfnamefont{J.}~\bibnamefont{Gal\'an}},
  \bibinfo{author}{\bibfnamefont{J.~P.} \bibnamefont{Kaminski}},
  \bibinfo{author}{\bibfnamefont{K.~L.} \bibnamefont{Campman}},
  \bibinfo{author}{\bibfnamefont{A.~C.} \bibnamefont{Gossard}},
  \bibinfo{author}{\bibfnamefont{U.}~\bibnamefont{Bhattacharya}},
  \bibnamefont{and} \bibinfo{author}{\bibfnamefont{M.~J.~W.}
  \bibnamefont{Rodwell}}, \bibinfo{journal}{Phys. Rev. Lett.}
  \textbf{\bibinfo{volume}{75}}, \bibinfo{pages}{4098}
  (\bibinfo{year}{1995}{\natexlab{a}}).

\bibitem[{\citenamefont{Keay et~al.}(1995{\natexlab{b}})\citenamefont{Keay,
  Zeuner, Allen, Maranowski, Gossard, Bhattacharya, and Rodwell}}]{Keay1995b}
\bibinfo{author}{\bibfnamefont{B.~J.} \bibnamefont{Keay}},
  \bibinfo{author}{\bibfnamefont{S.}~\bibnamefont{Zeuner}},
  \bibinfo{author}{\bibfnamefont{S.~J.} \bibnamefont{Allen}},
  \bibinfo{author}{\bibfnamefont{K.~D.} \bibnamefont{Maranowski}},
  \bibinfo{author}{\bibfnamefont{A.~C.} \bibnamefont{Gossard}},
  \bibinfo{author}{\bibfnamefont{U.}~\bibnamefont{Bhattacharya}},
  \bibnamefont{and} \bibinfo{author}{\bibfnamefont{M.~J.~W.}
  \bibnamefont{Rodwell}}, \bibinfo{journal}{Phys. Rev. Lett.}
  \textbf{\bibinfo{volume}{75}}, \bibinfo{pages}{4102}
  (\bibinfo{year}{1995}{\natexlab{b}}).

\bibitem[{\citenamefont{Kouwenhoven et~al.}(1994)\citenamefont{Kouwenhoven,
  Jauhar, Orenstein, McEuen, Nagamune, Motohisa, and Sakaki}}]{Kouwenhoven1994}
\bibinfo{author}{\bibfnamefont{L.~P.} \bibnamefont{Kouwenhoven}},
  \bibinfo{author}{\bibfnamefont{S.}~\bibnamefont{Jauhar}},
  \bibinfo{author}{\bibfnamefont{J.}~\bibnamefont{Orenstein}},
  \bibinfo{author}{\bibfnamefont{P.~L.} \bibnamefont{McEuen}},
  \bibinfo{author}{\bibfnamefont{Y.}~\bibnamefont{Nagamune}},
  \bibinfo{author}{\bibfnamefont{J.}~\bibnamefont{Motohisa}}, \bibnamefont{and}
  \bibinfo{author}{\bibfnamefont{H.}~\bibnamefont{Sakaki}},
  \bibinfo{journal}{Phys. Rev. Lett.} \textbf{\bibinfo{volume}{73}},
  \bibinfo{pages}{3443} (\bibinfo{year}{1994}).

\bibitem[{\citenamefont{Oosterkamp et~al.}(1997)\citenamefont{Oosterkamp,
  Kouwenhoven, Koolen, van~der Vaart, and Harmans}}]{Oosterkamp1997}
\bibinfo{author}{\bibfnamefont{T.~H.} \bibnamefont{Oosterkamp}},
  \bibinfo{author}{\bibfnamefont{L.~P.} \bibnamefont{Kouwenhoven}},
  \bibinfo{author}{\bibfnamefont{A.~E.~A.} \bibnamefont{Koolen}},
  \bibinfo{author}{\bibfnamefont{N.~C.} \bibnamefont{van~der Vaart}},
  \bibnamefont{and} \bibinfo{author}{\bibfnamefont{C.~J. P.~M.}
  \bibnamefont{Harmans}}, \bibinfo{journal}{Phys. Rev. Lett.}
  \textbf{\bibinfo{volume}{78}}, \bibinfo{pages}{1536} (\bibinfo{year}{1997}).

\bibitem[{\citenamefont{Eckardt et~al.}(2005)\citenamefont{Eckardt, Weiss, and
  Holthaus}}]{Eckardt2005}
\bibinfo{author}{\bibfnamefont{A.}~\bibnamefont{Eckardt}},
  \bibinfo{author}{\bibfnamefont{C.}~\bibnamefont{Weiss}}, \bibnamefont{and}
  \bibinfo{author}{\bibfnamefont{M.}~\bibnamefont{Holthaus}},
  \bibinfo{journal}{Phys. Rev. Lett.} \textbf{\bibinfo{volume}{95}},
  \bibinfo{pages}{260404} (\bibinfo{year}{2005}).

\bibitem[{\citenamefont{Kolovsky and Korsch}(2010)}]{Kolovsky2010}
\bibinfo{author}{\bibfnamefont{A.~R.} \bibnamefont{Kolovsky}} \bibnamefont{and}
  \bibinfo{author}{\bibfnamefont{H.~J.} \bibnamefont{Korsch}},
  \bibinfo{journal}{J. Sib. Fed. Un.: Math, Phys} \textbf{\bibinfo{volume}{3}},
  \bibinfo{pages}{211} (\bibinfo{year}{2010}).

\bibitem[{\citenamefont{Sias et~al.}(2008)\citenamefont{Sias, Lignier, Singh,
  Zenesini, Ciampini, Morsch, and Arimondo}}]{Sias2008}
\bibinfo{author}{\bibfnamefont{C.}~\bibnamefont{Sias}},
  \bibinfo{author}{\bibfnamefont{H.}~\bibnamefont{Lignier}},
  \bibinfo{author}{\bibfnamefont{Y.~P.} \bibnamefont{Singh}},
  \bibinfo{author}{\bibfnamefont{A.}~\bibnamefont{Zenesini}},
  \bibinfo{author}{\bibfnamefont{D.}~\bibnamefont{Ciampini}},
  \bibinfo{author}{\bibfnamefont{O.}~\bibnamefont{Morsch}}, \bibnamefont{and}
  \bibinfo{author}{\bibfnamefont{E.}~\bibnamefont{Arimondo}},
  \bibinfo{journal}{Phys. Rev. Lett.} \textbf{\bibinfo{volume}{100}},
  \bibinfo{eid}{040404} (\bibinfo{year}{2008}).

\bibitem[{\citenamefont{Ivanov et~al.}(2008)\citenamefont{Ivanov, Alberti,
  Schioppo, Ferrari, Artoni, Chiofalo, and Tino}}]{Ivanov2008}
\bibinfo{author}{\bibfnamefont{V.~V.} \bibnamefont{Ivanov}},
  \bibinfo{author}{\bibfnamefont{A.}~\bibnamefont{Alberti}},
  \bibinfo{author}{\bibfnamefont{M.}~\bibnamefont{Schioppo}},
  \bibinfo{author}{\bibfnamefont{G.}~\bibnamefont{Ferrari}},
  \bibinfo{author}{\bibfnamefont{M.}~\bibnamefont{Artoni}},
  \bibinfo{author}{\bibfnamefont{M.~L.} \bibnamefont{Chiofalo}},
  \bibnamefont{and} \bibinfo{author}{\bibfnamefont{G.~M.} \bibnamefont{Tino}},
  \bibinfo{journal}{Phys. Rev. Lett.} \textbf{\bibinfo{volume}{100}},
  \bibinfo{eid}{043602} (\bibinfo{year}{2008}).

\bibitem[{\citenamefont{Alberti et~al.}(2009)\citenamefont{Alberti, Ivanov,
  Tino, and Ferrari}}]{Alberti2009}
\bibinfo{author}{\bibfnamefont{A.}~\bibnamefont{Alberti}},
  \bibinfo{author}{\bibfnamefont{V.}~\bibnamefont{Ivanov}},
  \bibinfo{author}{\bibfnamefont{G.}~\bibnamefont{Tino}}, \bibnamefont{and}
  \bibinfo{author}{\bibfnamefont{G.}~\bibnamefont{Ferrari}},
  \bibinfo{journal}{Nat. Phys.} \textbf{\bibinfo{volume}{5}},
  \bibinfo{pages}{547} (\bibinfo{year}{2009}).

\bibitem[{\citenamefont{Haller et~al.}(2010)\citenamefont{Haller, Hart, Mark,
  Danzl, Reichs\"ollner, and N\"agerl}}]{Haller2010}
\bibinfo{author}{\bibfnamefont{E.}~\bibnamefont{Haller}},
  \bibinfo{author}{\bibfnamefont{R.}~\bibnamefont{Hart}},
  \bibinfo{author}{\bibfnamefont{M.~J.} \bibnamefont{Mark}},
  \bibinfo{author}{\bibfnamefont{J.}~\bibnamefont{Danzl}},
  \bibinfo{author}{\bibfnamefont{L.}~\bibnamefont{Reichs\"ollner}},
  \bibnamefont{and} \bibinfo{author}{\bibfnamefont{H.}~\bibnamefont{N\"agerl}},
  \bibinfo{journal}{Phys. Rev. Lett.} \textbf{\bibinfo{volume}{104}},
  \bibinfo{pages}{200403} (\bibinfo{year}{2010}).

\bibitem[{\citenamefont{Thommen et~al.}(2002)\citenamefont{Thommen, Garreau,
  and Zehnle}}]{Thommen2002}
\bibinfo{author}{\bibfnamefont{Q.}~\bibnamefont{Thommen}},
  \bibinfo{author}{\bibfnamefont{J.~C.} \bibnamefont{Garreau}},
  \bibnamefont{and} \bibinfo{author}{\bibfnamefont{V.}~\bibnamefont{Zehnle}},
  \bibinfo{journal}{Phys. Rev. A} \textbf{\bibinfo{volume}{65}},
  \bibinfo{eid}{053406} (\bibinfo{year}{2002}).

\bibitem[{\citenamefont{Eckardt and Holthaus}(2008)}]{Eckardt2008}
\bibinfo{author}{\bibfnamefont{A.}~\bibnamefont{Eckardt}} \bibnamefont{and}
  \bibinfo{author}{\bibfnamefont{M.}~\bibnamefont{Holthaus}},
  \bibinfo{journal}{J. Phys: Conf. Ser.} \textbf{\bibinfo{volume}{99}},
  \bibinfo{pages}{012007} (\bibinfo{year}{2008}).

\bibitem[{\citenamefont{Lignier et~al.}(2007)\citenamefont{Lignier, Sias,
  Ciampini, Singh, Zenesini, Morsch, and Arimondo}}]{Lignier2007}
\bibinfo{author}{\bibfnamefont{H.}~\bibnamefont{Lignier}},
  \bibinfo{author}{\bibfnamefont{C.}~\bibnamefont{Sias}},
  \bibinfo{author}{\bibfnamefont{D.}~\bibnamefont{Ciampini}},
  \bibinfo{author}{\bibfnamefont{Y.}~\bibnamefont{Singh}},
  \bibinfo{author}{\bibfnamefont{A.}~\bibnamefont{Zenesini}},
  \bibinfo{author}{\bibfnamefont{O.}~\bibnamefont{Morsch}}, \bibnamefont{and}
  \bibinfo{author}{\bibfnamefont{E.}~\bibnamefont{Arimondo}},
  \bibinfo{journal}{Phys. Rev. Lett.} \textbf{\bibinfo{volume}{99}},
  \bibinfo{eid}{220403} (\bibinfo{year}{2007}).

\bibitem[{\citenamefont{Bharucha et~al.}(1997)\citenamefont{Bharucha, Madison,
  Morrow, Wilkinson, Sundaram, and Raizen}}]{Bharucha1997}
\bibinfo{author}{\bibfnamefont{C.~F.} \bibnamefont{Bharucha}},
  \bibinfo{author}{\bibfnamefont{K.~W.} \bibnamefont{Madison}},
  \bibinfo{author}{\bibfnamefont{P.~R.} \bibnamefont{Morrow}},
  \bibinfo{author}{\bibfnamefont{S.~R.} \bibnamefont{Wilkinson}},
  \bibinfo{author}{\bibfnamefont{B.}~\bibnamefont{Sundaram}}, \bibnamefont{and}
  \bibinfo{author}{\bibfnamefont{M.~G.} \bibnamefont{Raizen}},
  \bibinfo{journal}{Phys. Rev. A} \textbf{\bibinfo{volume}{55}},
  \bibinfo{pages}{R857} (\bibinfo{year}{1997}).

\bibitem[{\citenamefont{Wimberger et~al.}(2005)\citenamefont{Wimberger,
  Mannella, Morsch, Arimondo, Kolovsky, and Buchleitner}}]{Wimberger2005}
\bibinfo{author}{\bibfnamefont{S.}~\bibnamefont{Wimberger}},
  \bibinfo{author}{\bibfnamefont{R.}~\bibnamefont{Mannella}},
  \bibinfo{author}{\bibfnamefont{O.}~\bibnamefont{Morsch}},
  \bibinfo{author}{\bibfnamefont{E.}~\bibnamefont{Arimondo}},
  \bibinfo{author}{\bibfnamefont{A.~R.} \bibnamefont{Kolovsky}},
  \bibnamefont{and}
  \bibinfo{author}{\bibfnamefont{A.}~\bibnamefont{Buchleitner}},
  \bibinfo{journal}{Phys. Rev. A} \textbf{\bibinfo{volume}{72}},
  \bibinfo{pages}{063610} (\bibinfo{year}{2005}).

\bibitem[{\citenamefont{Avron}(1982)}]{Avron1982}
\bibinfo{author}{\bibfnamefont{J.~E.} \bibnamefont{Avron}},
  \bibinfo{journal}{Annals of Physics} \textbf{\bibinfo{volume}{143}},
  \bibinfo{pages}{33} (\bibinfo{year}{1982}).

\bibitem[{\citenamefont{Keck et~al.}(2003)\citenamefont{Keck, Korsch, and
  Mossmann}}]{Keck2003}
\bibinfo{author}{\bibfnamefont{F.}~\bibnamefont{Keck}},
  \bibinfo{author}{\bibfnamefont{H.~J.} \bibnamefont{Korsch}},
  \bibnamefont{and} \bibinfo{author}{\bibfnamefont{S.}~\bibnamefont{Mossmann}},
  \bibinfo{journal}{J. Phys. A: Mathem. Gen.} \textbf{\bibinfo{volume}{36}},
  \bibinfo{pages}{2125} (\bibinfo{year}{2003}).

\bibitem[{\citenamefont{Gl\"uck et~al.}(1999)\citenamefont{Gl\"uck, Kolovsky,
  and Korsch}}]{Glueck1999}
\bibinfo{author}{\bibfnamefont{M.}~\bibnamefont{Gl\"uck}},
  \bibinfo{author}{\bibfnamefont{A.~R.} \bibnamefont{Kolovsky}},
  \bibnamefont{and} \bibinfo{author}{\bibfnamefont{H.~J.}
  \bibnamefont{Korsch}}, \bibinfo{journal}{Phys. Rev. Lett.}
  \textbf{\bibinfo{volume}{83}}, \bibinfo{pages}{891} (\bibinfo{year}{1999}).

\bibitem[{\citenamefont{Wimberger et~al.}(2006)\citenamefont{Wimberger,
  Schlagheck, and Mannella}}]{Wimberger2006}
\bibinfo{author}{\bibfnamefont{S.}~\bibnamefont{Wimberger}},
  \bibinfo{author}{\bibfnamefont{P.}~\bibnamefont{Schlagheck}},
  \bibnamefont{and} \bibinfo{author}{\bibfnamefont{R.}~\bibnamefont{Mannella}},
  \bibinfo{journal}{J. Phys. B: At. Molec. Opt. Phys.}
  \textbf{\bibinfo{volume}{39}}, \bibinfo{pages}{729} (\bibinfo{year}{2006}).

\bibitem[{\citenamefont{Schlagheck and Wimberger}(2007)}]{Schlagheck2007}
\bibinfo{author}{\bibfnamefont{P.}~\bibnamefont{Schlagheck}} \bibnamefont{and}
  \bibinfo{author}{\bibfnamefont{S.}~\bibnamefont{Wimberger}},
  \bibinfo{journal}{Applied Physics B: Lasers and Optics}
  \textbf{\bibinfo{volume}{86}}, \bibinfo{pages}{385} (\bibinfo{year}{2007}).

\bibitem[{\citenamefont{Witthaut et~al.}(2007)\citenamefont{Witthaut, Graefe,
  Wimberger, and Korsch}}]{Witthaut2007}
\bibinfo{author}{\bibfnamefont{D.}~\bibnamefont{Witthaut}},
  \bibinfo{author}{\bibfnamefont{E.~M.} \bibnamefont{Graefe}},
  \bibinfo{author}{\bibfnamefont{S.}~\bibnamefont{Wimberger}},
  \bibnamefont{and} \bibinfo{author}{\bibfnamefont{H.~J.}
  \bibnamefont{Korsch}}, \bibinfo{journal}{Phys. Rev. A}
  \textbf{\bibinfo{volume}{75}}, \bibinfo{eid}{013617} (\bibinfo{year}{2007}).

\bibitem[{\citenamefont{Buchleitner and Kolovsky}(2003)}]{Buchleitner2003}
\bibinfo{author}{\bibfnamefont{A.}~\bibnamefont{Buchleitner}} \bibnamefont{and}
  \bibinfo{author}{\bibfnamefont{A.~R.} \bibnamefont{Kolovsky}},
  \bibinfo{journal}{Phys. Rev. Lett.} \textbf{\bibinfo{volume}{91}},
  \bibinfo{pages}{253002} (\bibinfo{year}{2003}).

\bibitem[{\citenamefont{Thommen et~al.}(2003)\citenamefont{Thommen, Garreau,
  and Zehnl\'e}}]{Thommen2003}
\bibinfo{author}{\bibfnamefont{Q.}~\bibnamefont{Thommen}},
  \bibinfo{author}{\bibfnamefont{J.~C.} \bibnamefont{Garreau}},
  \bibnamefont{and} \bibinfo{author}{\bibfnamefont{V.}~\bibnamefont{Zehnl\'e}},
  \bibinfo{journal}{Phys. Rev. Lett.} \textbf{\bibinfo{volume}{91}},
  \bibinfo{pages}{210405} (\bibinfo{year}{2003}).

\bibitem[{\citenamefont{Tomadin et~al.}(2007)\citenamefont{Tomadin, Mannella,
  and Wimberger}}]{Tomadin2007}
\bibinfo{author}{\bibfnamefont{A.}~\bibnamefont{Tomadin}},
  \bibinfo{author}{\bibfnamefont{R.}~\bibnamefont{Mannella}}, \bibnamefont{and}
  \bibinfo{author}{\bibfnamefont{S.}~\bibnamefont{Wimberger}},
  \bibinfo{journal}{Phys. Rev. Lett.} \textbf{\bibinfo{volume}{98}},
  \bibinfo{eid}{130402} (\bibinfo{year}{2007}).

\bibitem[{\citenamefont{Tomadin et~al.}(2008)\citenamefont{Tomadin, Mannella,
  and Wimberger}}]{Tomadin2008}
\bibinfo{author}{\bibfnamefont{A.}~\bibnamefont{Tomadin}},
  \bibinfo{author}{\bibfnamefont{R.}~\bibnamefont{Mannella}}, \bibnamefont{and}
  \bibinfo{author}{\bibfnamefont{S.}~\bibnamefont{Wimberger}},
  \bibinfo{journal}{Phys. Rev. A} \textbf{\bibinfo{volume}{77}},
  \bibinfo{eid}{013606} (\bibinfo{year}{2008}).

\bibitem[{\citenamefont{Buonsante and Wimberger}(2008)}]{Buonsante2008}
\bibinfo{author}{\bibfnamefont{P.}~\bibnamefont{Buonsante}} \bibnamefont{and}
  \bibinfo{author}{\bibfnamefont{S.}~\bibnamefont{Wimberger}},
  \bibinfo{journal}{Phys. Rev. A} \textbf{\bibinfo{volume}{77}},
  \bibinfo{pages}{041606(R)} (\bibinfo{year}{2008}).

\bibitem[{\citenamefont{Wilkinson et~al.}(1997)\citenamefont{Wilkinson,
  Bharucha, Fischer, Madison, Morrow, Niu, Sundaram, and
  Raizen}}]{Wilkinson1997}
\bibinfo{author}{\bibfnamefont{S.~R.} \bibnamefont{Wilkinson}},
  \bibinfo{author}{\bibfnamefont{C.~F.} \bibnamefont{Bharucha}},
  \bibinfo{author}{\bibfnamefont{M.~C.} \bibnamefont{Fischer}},
  \bibinfo{author}{\bibfnamefont{K.~W.} \bibnamefont{Madison}},
  \bibinfo{author}{\bibfnamefont{P.~R.} \bibnamefont{Morrow}},
  \bibinfo{author}{\bibfnamefont{Q.}~\bibnamefont{Niu}},
  \bibinfo{author}{\bibfnamefont{B.}~\bibnamefont{Sundaram}}, \bibnamefont{and}
  \bibinfo{author}{\bibfnamefont{M.~G.} \bibnamefont{Raizen}},
  \bibinfo{journal}{Nature} \textbf{\bibinfo{volume}{387}},
  \bibinfo{pages}{575} (\bibinfo{year}{1997}).

\bibitem[{\citenamefont{Carr et~al.}(2005)\citenamefont{Carr, Holland, and
  Malomed}}]{Carr2005}
\bibinfo{author}{\bibfnamefont{L.~D.} \bibnamefont{Carr}},
  \bibinfo{author}{\bibfnamefont{M.~J.} \bibnamefont{Holland}},
  \bibnamefont{and} \bibinfo{author}{\bibfnamefont{B.~A.}
  \bibnamefont{Malomed}}, \bibinfo{journal}{J. Phys. B: At. Molec. Opt. Phys.}
  \textbf{\bibinfo{volume}{38}}, \bibinfo{pages}{3217} (\bibinfo{year}{2005}).

\bibitem[{\citenamefont{Wimberger et~al.}(2007)\citenamefont{Wimberger,
  Ciampini, Morsch, Mannella, and Arimondo}}]{Wimberger2007}
\bibinfo{author}{\bibfnamefont{S.}~\bibnamefont{Wimberger}},
  \bibinfo{author}{\bibfnamefont{D.}~\bibnamefont{Ciampini}},
  \bibinfo{author}{\bibfnamefont{O.}~\bibnamefont{Morsch}},
  \bibinfo{author}{\bibfnamefont{R.}~\bibnamefont{Mannella}}, \bibnamefont{and}
  \bibinfo{author}{\bibfnamefont{E.}~\bibnamefont{Arimondo}},
  \bibinfo{journal}{J. Phys.: Conference Series} \textbf{\bibinfo{volume}{67}},
  \bibinfo{pages}{012060} (\bibinfo{year}{2007}).

\bibitem[{\citenamefont{Morsch et~al.}(2001)\citenamefont{Morsch, M\"uller,
  Cristiani, Ciampini, and Arimondo}}]{Morsch2001}
\bibinfo{author}{\bibfnamefont{O.}~\bibnamefont{Morsch}},
  \bibinfo{author}{\bibfnamefont{J.~H.} \bibnamefont{M\"uller}},
  \bibinfo{author}{\bibfnamefont{M.}~\bibnamefont{Cristiani}},
  \bibinfo{author}{\bibfnamefont{D.}~\bibnamefont{Ciampini}}, \bibnamefont{and}
  \bibinfo{author}{\bibfnamefont{E.}~\bibnamefont{Arimondo}},
  \bibinfo{journal}{Phys. Rev. Lett.} \textbf{\bibinfo{volume}{87}},
  \bibinfo{pages}{140402} (\bibinfo{year}{2001}).

\bibitem[{\citenamefont{Jona-Lasinio et~al.}(2003)\citenamefont{Jona-Lasinio,
  Morsch, Cristiani, Malossi, M\"uller, Courtade, Anderlini, and
  Arimondo}}]{JonaLasinio2003}
\bibinfo{author}{\bibfnamefont{M.}~\bibnamefont{Jona-Lasinio}},
  \bibinfo{author}{\bibfnamefont{O.}~\bibnamefont{Morsch}},
  \bibinfo{author}{\bibfnamefont{M.}~\bibnamefont{Cristiani}},
  \bibinfo{author}{\bibfnamefont{N.}~\bibnamefont{Malossi}},
  \bibinfo{author}{\bibfnamefont{J.~H.} \bibnamefont{M\"uller}},
  \bibinfo{author}{\bibfnamefont{E.}~\bibnamefont{Courtade}},
  \bibinfo{author}{\bibfnamefont{M.}~\bibnamefont{Anderlini}},
  \bibnamefont{and} \bibinfo{author}{\bibfnamefont{E.}~\bibnamefont{Arimondo}},
  \bibinfo{journal}{Phys. Rev. Lett.} \textbf{\bibinfo{volume}{91}},
  \bibinfo{pages}{230406} (\bibinfo{year}{2003}).

\bibitem[{\citenamefont{Choi and Niu}(1999)}]{ChoiNiu1999}
\bibinfo{author}{\bibfnamefont{D.-I.} \bibnamefont{Choi}} \bibnamefont{and}
  \bibinfo{author}{\bibfnamefont{Q.}~\bibnamefont{Niu}},
  \bibinfo{journal}{Phys. Rev. Lett.} \textbf{\bibinfo{volume}{82}},
  \bibinfo{pages}{2022} (\bibinfo{year}{1999}).

\bibitem[{\citenamefont{Krimer et~al.}(2009)\citenamefont{Krimer, Khomeriki,
  and Flach}}]{Krimer2009}
\bibinfo{author}{\bibfnamefont{D.~O.} \bibnamefont{Krimer}},
  \bibinfo{author}{\bibfnamefont{R.}~\bibnamefont{Khomeriki}},
  \bibnamefont{and} \bibinfo{author}{\bibfnamefont{S.}~\bibnamefont{Flach}},
  \bibinfo{journal}{Phys. Rev. E} \textbf{\bibinfo{volume}{80}},
  \bibinfo{eid}{036201} (\bibinfo{year}{2009}).

\bibitem[{\citenamefont{K\"uhner and Monien}(1998)}]{Kuehner1998}
\bibinfo{author}{\bibfnamefont{T.~D.} \bibnamefont{K\"uhner}} \bibnamefont{and}
  \bibinfo{author}{\bibfnamefont{H.}~\bibnamefont{Monien}},
  \bibinfo{journal}{Phys. Rev. B} \textbf{\bibinfo{volume}{58}},
  \bibinfo{pages}{R14741} (\bibinfo{year}{1998}).

\bibitem[{\citenamefont{Duine and Stoof}(2003)}]{Duine2003}
\bibinfo{author}{\bibfnamefont{R.~A.} \bibnamefont{Duine}} \bibnamefont{and}
  \bibinfo{author}{\bibfnamefont{H.~T.~C.} \bibnamefont{Stoof}},
  \bibinfo{journal}{Phys. Rev. Lett.} \textbf{\bibinfo{volume}{91}},
  \bibinfo{pages}{150405} (\bibinfo{year}{2003}).

\bibitem[{\citenamefont{Kolovsky and Buchleitner}(2003)}]{Kolovsky2003}
\bibinfo{author}{\bibfnamefont{A.~R.} \bibnamefont{Kolovsky}} \bibnamefont{and}
  \bibinfo{author}{\bibfnamefont{A.}~\bibnamefont{Buchleitner}},
  \bibinfo{journal}{Phys. Rev. E} \textbf{\bibinfo{volume}{68}},
  \bibinfo{pages}{056213} (\bibinfo{year}{2003}).

\bibitem[{\citenamefont{Pl\"otz et~al.}(2010)\citenamefont{Pl\"otz,
  Madro{\~n}ero, and Wimberger}}]{Ploetz2010}
\bibinfo{author}{\bibfnamefont{P.}~\bibnamefont{Pl\"otz}},
  \bibinfo{author}{\bibfnamefont{J.}~\bibnamefont{Madro{\~n}ero}},
  \bibnamefont{and}
  \bibinfo{author}{\bibfnamefont{S.}~\bibnamefont{Wimberger}},
  \bibinfo{journal}{J. Phys. B: At. Mol. Opt. Phys.}
  \textbf{\bibinfo{volume}{43}}, \bibinfo{pages}{081001}
  (\bibinfo{year}{2010}).

\bibitem[{\citenamefont{Beenakker}(1997)}]{Beenakker1997}
\bibinfo{author}{\bibfnamefont{C.~W.~J.} \bibnamefont{Beenakker}},
  \bibinfo{journal}{Rev. Mod. Phys.} \textbf{\bibinfo{volume}{69}},
  \bibinfo{pages}{731} (\bibinfo{year}{1997}).

\bibitem[{\citenamefont{Kottos}(2005)}]{Kottos2005}
\bibinfo{author}{\bibfnamefont{T.}~\bibnamefont{Kottos}}, \bibinfo{journal}{J.
  Phys. A: Mathematical and General} \textbf{\bibinfo{volume}{38}},
  \bibinfo{pages}{10761} (\bibinfo{year}{2005}).

\bibitem[{\citenamefont{Nakamura et~al.}(2001)\citenamefont{Nakamura, Pashkin,
  and Tsai}}]{Nakamura2001}
\bibinfo{author}{\bibfnamefont{Y.}~\bibnamefont{Nakamura}},
  \bibinfo{author}{\bibfnamefont{Y.~A.} \bibnamefont{Pashkin}},
  \bibnamefont{and} \bibinfo{author}{\bibfnamefont{J.~S.} \bibnamefont{Tsai}},
  \bibinfo{journal}{Phys. Rev. Lett.} \textbf{\bibinfo{volume}{87}},
  \bibinfo{pages}{246601} (\bibinfo{year}{2001}).

\bibitem[{\citenamefont{Meystre and Sargent}(2007)}]{Meystre2007}
\bibinfo{author}{\bibfnamefont{P.}~\bibnamefont{Meystre}} \bibnamefont{and}
  \bibinfo{author}{\bibfnamefont{M.~I.} \bibnamefont{Sargent}},
  \emph{\bibinfo{title}{Elements of Quantum Optics}}
  (\bibinfo{publisher}{Springer Verlag}, \bibinfo{address}{Heidelberg},
  \bibinfo{year}{2007}).

\bibitem[{\citenamefont{Anderlini et~al.}(2006)\citenamefont{Anderlini,
  Sebby-Strabley, Kruse, Porto, and Phillips}}]{Anderlini2006}
\bibinfo{author}{\bibfnamefont{M.}~\bibnamefont{Anderlini}},
  \bibinfo{author}{\bibfnamefont{J.}~\bibnamefont{Sebby-Strabley}},
  \bibinfo{author}{\bibfnamefont{J.}~\bibnamefont{Kruse}},
  \bibinfo{author}{\bibfnamefont{J.~V.} \bibnamefont{Porto}}, \bibnamefont{and}
  \bibinfo{author}{\bibfnamefont{W.}~\bibnamefont{Phillips}},
  \bibinfo{journal}{J. Phys. B: At. Mol. and Opt. Phys.}
  \textbf{\bibinfo{volume}{39}}, \bibinfo{pages}{S199} (\bibinfo{year}{2006}).

\bibitem[{\citenamefont{Sebby-Strabley
  et~al.}(2007)\citenamefont{Sebby-Strabley, Brown, Anderlini, Lee, Phillips,
  Porto, and Johnson}}]{SebbyStrabley2007}
\bibinfo{author}{\bibfnamefont{J.}~\bibnamefont{Sebby-Strabley}},
  \bibinfo{author}{\bibfnamefont{B.~L.} \bibnamefont{Brown}},
  \bibinfo{author}{\bibfnamefont{M.}~\bibnamefont{Anderlini}},
  \bibinfo{author}{\bibfnamefont{P.~J.} \bibnamefont{Lee}},
  \bibinfo{author}{\bibfnamefont{W.~D.} \bibnamefont{Phillips}},
  \bibinfo{author}{\bibfnamefont{J.~V.} \bibnamefont{Porto}}, \bibnamefont{and}
  \bibinfo{author}{\bibfnamefont{P.~R.} \bibnamefont{Johnson}},
  \bibinfo{journal}{Phys. Rev. Lett.} \textbf{\bibinfo{volume}{98}},
  \bibinfo{eid}{200405} (\bibinfo{year}{2007}).

\bibitem[{\citenamefont{Will et~al.}(2010)\citenamefont{Will, Best, Schneider,
  HackermŸller, L\"uhmann, and Bloch}}]{Will2010}
\bibinfo{author}{\bibfnamefont{S.}~\bibnamefont{Will}},
  \bibinfo{author}{\bibfnamefont{T.}~\bibnamefont{Best}},
  \bibinfo{author}{\bibfnamefont{U.}~\bibnamefont{Schneider}},
  \bibinfo{author}{\bibfnamefont{L.}~\bibnamefont{HackermŸller}},
  \bibinfo{author}{\bibfnamefont{D.-S.} \bibnamefont{L\"uhmann}},
  \bibnamefont{and} \bibinfo{author}{\bibfnamefont{I.}~\bibnamefont{Bloch}},
  \bibinfo{journal}{Nature} \textbf{\bibinfo{volume}{465}},
  \bibinfo{pages}{197} (\bibinfo{year}{2010}).

\bibitem[{\citenamefont{Ueda and Leggett}(1998{\natexlab{a}})}]{Ueda1998}
\bibinfo{author}{\bibfnamefont{M.}~\bibnamefont{Ueda}} \bibnamefont{and}
  \bibinfo{author}{\bibfnamefont{A.~J.} \bibnamefont{Leggett}},
  \bibinfo{journal}{Phys. Rev. Lett.} \textbf{\bibinfo{volume}{80}},
  \bibinfo{pages}{1576} (\bibinfo{year}{1998}{\natexlab{a}}).

\bibitem[{\citenamefont{Ueda and Leggett}(1998{\natexlab{b}})}]{UedaReply1998}
\bibinfo{author}{\bibfnamefont{M.}~\bibnamefont{Ueda}} \bibnamefont{and}
  \bibinfo{author}{\bibfnamefont{A.~J.} \bibnamefont{Leggett}},
  \bibinfo{journal}{Phys. Rev. Lett.} \textbf{\bibinfo{volume}{81}},
  \bibinfo{pages}{1343} (\bibinfo{year}{1998}{\natexlab{b}}).

\end{thebibliography}

\end{document}